\documentclass[nohyper,12pt,letterpaper]{JHEP3}

\usepackage{epsfig}
\usepackage[latin1]{inputenc}
\usepackage{bbm,amsfonts}
\usepackage{graphicx}
\usepackage{amssymb,amsmath}

\usepackage{latexsym, stmaryrd}
\usepackage{dsfont}

\newcommand{\graph}[2]{\vcenter{\hbox{\includegraphics[height=#2ex]{#1}}}}
\newcommand{\di}{\text{d}}
\renewcommand{\d}{\text{d}}
\renewcommand{\k}{|2k\rangle}

\author{Andrea Mauri$^{\dag\hash}$,
  Alberto Santambrogio$^{\hash}$ and
  Stefano Scoleri$^{\dag \hash}$ \\\\
  $^\dag$Dipartimento di Fisica dell'Universit\`a degli studi di Milano, via Celoria 16, I-20133 Milano, Italy\\\\
  $^\hash$ INFN, Sezione di Milano, via Celoria 16, I-20133 Milano, Italy
  \qquad\\\\
  E-mail: \email{andrea.mauri@mi.infn.it, alberto.santambrogio@mi.infn.it, stefano.scoleri@unimi.it }\\}

\abstract{
We study the planar asymptotic dilatation operator of ABJM theory
in the $SU(2)\times SU(2)$ sector up to eight loops. Combining
Bethe Ansatz techniques and $\mathcal{N}=2$ superspace methods, we
are able to fix all the coefficients appearing in the
maximal-reshuffling terms. In particular, we can directly compute
from Feynman diagrams the leading order coefficient
$\beta_{2,3}^{(6)}$ of the dressing phase and find an agreement with the relation
conjectured by  Gromov and Vieira  between the ABJM and $\mathcal{N}=4$ SYM phase factor.}

\preprint{January 2013 \\ IFUM-1006-FT}

\title{The Leading-Order Dressing Phase in ABJM Theory}

\keywords{AdS/CFT, Chern--Simons matter theories, integrability, dressing phase}

\begin{document}

\section{Introduction}
In the last few years remarkable mathematical structures were
shown to emerge in the analysis of  supersymmetric
Chern-Simons-matter theories in three dimensions.  In this class
of theories a distinguished role is played by the  $\mathcal{N}=6$
ABJM model \cite{Aharony:2008ug} which is  a $U(N)_k \times
U(N)_{-k}$ superconformal gauge theory with Chern-Simons level
$k$.  Indeed, in the large $N$ limit the  ABJM theory has been
conjectured to be the AdS/CFT  dual description of M-theory on an
$AdS_4 \times S_7/Z_k$ background and, for $k \ll N \ll k^5$, of a
type IIA string theory on $AdS_4 \times CP_3$.  For this reason,
soon after its discovery the ABJM model has quickly become the
ideal three-dimensional playground to study AdS/CFT as much as
$\mathcal{N}=4$ SYM has been in the four-dimensional case.

Quite surprisingly, the ABJM model seems to share a number of
remarkable properties with $\mathcal{N}=4$ SYM theory  even if the
two theories are {\it a priori} different in nature.   One of the
common features is provided by the fact that also in the ABJM
case, in the limit of large $N$ and $k$ with $\lambda=N/k$ kept
fixed, integrable structures naturally show up (for a review see
\cite{Beisert:2010jr}).

At first it was found in \cite{Minahan:2008hf, Bak:2008cp} that, at the
two-loop order and in the $SU(4)$ flavor sector,  the anomalous
dimensions of composite operators could be mapped to the energy
spectrum of an integrable Hamiltonian acting on an alternating
fundamental-antifundamental  spin-chain.

The two-loop  analysis was then extended to the full theory in
\cite{Minahan:2009te, Zwiebel:2009vb} by the introduction of an
$OSp(2,2|6)$ chain. The parity breaking ABJ model
\cite{Aharony:2008gk} was also studied at two loops in
\cite{Zwiebel:2009vb,Bak:2008vd}, where it was found to be
integrable at the given order.

Soon afterwards, paralleling the progresses done in the
four-dimensional case,  a set of all-loop  Bethe equations for
the asymptotic spectrum of the full ABJM theory was proposed in
\cite{Gromov:2008qe}. The Bethe equations nicely interpolated
between  the weak coupling results and the coset string
construction at strong coupling
\cite{Arutyunov:2008if,Stefanski:2008ik}, together with the
algebraic curve approach developed in \cite{Gromov:2008bz}.

One of the salient features of the Bethe equations introduced by Gromov 
and Vieira is that they are strongly constrained by the symmetries
of the theory. In fact only a pair of undetermined functions of
the coupling $\lambda$ are left open in the description of the
spectrum. One is the interpolating function $h(\lambda)$  which
relates the weak and strong coupling regimes of the single magnon
dispersion relation.  The other one is the dressing function
$\theta(\lambda)$, which  had to be introduced also in the
$\mathcal{N}=4$ SYM case \cite{Arutyunov:2004vx, Beisert:2005wv, Arutyunov:2006iu, Beisert:2006ib, Beisert:2006ez, Arutyunov:2009kf} 
 in order to have a proper
matching between the weak and strong coupling descriptions.  In
the ABJM case, the dressing phase plays an even more fundamental role 
since its presence has also been conjectured in \cite{Gromov:2008qe}
to give rise to the coupling between even- and odd-site excitations on the chain at
high-loop orders.

The  description of the asymptotic spectrum depicted in
\cite{Gromov:2008qe} has been subsequently checked at the
perturbative level beyond two-loops by direct Feynman diagrammatic
computations in \cite{Minahan:2009aq, Minahan:2009wg,
Leoni:2010tb, Bak:2009mq}. The four-loop dilatation operator has
been fully computed for both the ABJM and ABJ models by  using the
component formulation in \cite{Minahan:2009aq, Minahan:2009wg} and
the $\mathcal{N}=2$ superspace formalism in \cite{Leoni:2010tb}.
As a result,  the form of dilatation operator was found to be
compatible with the spectrum predicted by the Bethe equations and
moreover it was possible to fix the  next-to-leading order
coefficient in the weak coupling expansion of the function
$h(\lambda)$.

In \cite{Bak:2009tq} the analysis was pushed up to six loops.  At
this order, a full Feynman diagrammatic analysis looks very
complicated, even using superspace techniques. Nevertheless, in
\cite{Bak:2009tq} the expression of the dilatation operator could
be derived by computing  a suitable set of Feynman diagrams. For
further perturbative checks on the spectrum see
\cite{Papathanasiou:2009zm}.

Meanwhile, the internal  S-matrix approach to integrability was
developed in a number of papers \cite{Ahn:2008aa,
Ahn:2008tv, Ahn:2009zg, Ahn:2009tj, Ahn:2010eg, Hatsuda:2009pc}. An all loop
S-matrix has been found which is compatible with the all-loop Bethe
equations of \cite{Gromov:2008qe}. Moreover, the thermodinamic Bethe 
Ansatz and  Y-system framework has also been applied to the three dimensional 
case \cite{Gromov:2009tv, Bombardelli:2009xz, Gromov:2009at, LevkovichMaslyuk:2011ty}.

In the present paper we would like to provide further evidence for
the exactness of the conjectured integrable scenario by performing
a direct computation of the first non-vanishing coefficient of the
Bethe Ansatz dressing phase. We restrict ourselves to the  $SU(2)
\times SU(2)$ scalar sector of the theory, where the spin chain is
simply given by a pair of coupled $SU(2)$ spin chains.  In
\cite{Gromov:2008qe} it was mentioned that the dressing phase
starts contributing at eight loops, leading to the result that the
two  $SU(2)$  chains at odd and even sites are decoupled up to six
loops.  Indeed, it was explicitly shown in
\cite{Minahan:2009aq,Minahan:2009wg} and \cite{Leoni:2010tb} by
direct computations, that this is true at four-loop order: the
contributions to the dilatation operator of the diagrams that
could lead to interactions between the two types of magnons
cancel. Moreover, the results of \cite{Bak:2009tq} imply that this
is true also at six-loop order. This latter fact isn't so trivial
and has the consequence that the order $\lambda^4$ coefficient of
the dressing factor which can be a priori present at six loops,
actually vanishes. Therefore, the first non-trivial effects of the
presence of the dressing phase are to be found at eight-loop
order. With the aim to check the above picture, we analyze the form of
the eight-loop dilatation operator and extract the value of
the leading order coefficient of the dressing factor.  We make use
of a procedure inspired to the one used in \cite{Beisert:2007hz} for 
the computation of the analogous coefficient 
in $\mathcal{N}=4$ SYM.

The plan of the paper is the following. After reviewing in Section 2 the
aspects of integrability in the ABJM model we shall need,  in Section 3 we
introduce a procedure that is useful to constrain the form of
the dilatation operator as much as possible. This procedure is
based on symmetry arguments and on a matching  between the
entries of the diagonalized dilatation matrix with the Bethe
equations predictions for the energies of one and two-impurity
states.  We then use this procedure to write the maximal
reshuffling part of the dilatation operator in terms of a small
number of unknown parameters, one of which is the dressing phase.
In Section 4, we perform a direct Feynman diagram computation of
the maximal reshuffling diagrams to fix the unknown coefficients
including the dressing phase. We end up with a result which is in
complete agreement with the integrable picture conjectured in
\cite{Gromov:2008qe}.  As we shall discuss, our computation,
besides providing the explicit value for the leading coefficient
of the dressing factor, represents a non-trivial consistency check
of the Bethe equations, since, as mentioned before, the dressing
phase is also shown to play the structural role of coupling even
and odd excitations in the ABJM model. We give further comments on
our results in Section 5 while several technical aspects are
collected in the Appendices.

\section{The Bethe Ansatz in the $SU(2)\times SU(2)$ Sector}
\label{BA} In this section, we collect the integrability tools we
need for the construction of the dilatation operator. We refer to Appendix \ref{superspace} for the formulation of ABJM theory in terms of $\mathcal{N}=2$ superfields. 
Throughout the entire work, we restrict ourselves to the $SU(2)\times ~SU(2)$
sector, where operators are made out only of the chiral
superfields $Z^A$, $W_B$:
\begin{equation}\label{su2operators}
\mathcal{O}_{B_1\cdots B_L}^{A_1\cdots
A_L}=\text{Tr}[Z^{A_1}W_{B_1}\cdots Z^{A_L}W_{B_L}]\, ,
\end{equation}
with $A_j,B_j=1,2$. This sector is closed under renormalization at all loop-order.
In the spin-chain picture, the operators (\ref{su2operators}) are
mapped to states of a circular alternating
spin-chain, with the fields
$Z^A$ being interpreted as spins lying on odd sites and the $W_A$
as spins lying on even sites. We will refer to fields $Z^2$ and
$W_2$ as impurities and they correspond to ``spin down'' states.
The ground state of length $2L$ is chosen to be
\begin{equation}
 |\,0\,\rangle=\,\text{Tr}[Z^{1}W_{1}\cdots Z^{1}W_{1}]\, ,
\end{equation}
while the operators
\begin{equation}\begin{split}
 |2k\,\rangle=&\,\text{Tr}[Z^{1}W_{1}\cdots Z^{1}W_{2}\cdots Z^{1}W_{1}]\, , \\
|2k+1\,\rangle=&\,\text{Tr}[Z^{1}W_{1}\cdots Z^{2}W_{1}\cdots
Z^{1}W_{1}]
\end{split}\end{equation}
represent states with a single excitation on the site $2k$ and $2k+1$ respectively. There are two kinds of
magnon states, depending on their momentum $p$ being excited on
even or odd sites\footnote{In the sum, the identification
$2L+1\sim 1$ is understood.}:
\begin{equation}\label{magnons}\begin{split}
 |p\rangle_e&=\sum_{k=1}^Le^{ipk}\k\, ,\\
|p\rangle_o&=\sum_{k=1}^Le^{ipk}|2k+1\rangle\, .
\end{split}\end{equation}
The magnon dispersion relation is given by
\begin{equation}\label{dispersion}
 E(p)=\,\frac{1}{2}\,\left(\sqrt{1+16\,h^2(\lambda)\sin^2\frac{p}{2}}-1\right)\,
 ,
\end{equation}
where $h(\lambda)$ is the interpolating function,
which has the weak-coupling expansion \footnote{For discussions on the strong coupling expansion of $h(\lambda)$ see \cite{Grignani}-\cite{Beccaria:2012qd}.}
\begin{equation}
h^2(\lambda)=\sum_{k=1}^{\infty}\,h_{2k}\lambda^{2k}
\end{equation}
with $h_2=1$ and $h_4=-4\zeta(2)$ the only known coefficients
\cite{Minahan:2009aq,Minahan:2009wg,Leoni:2010tb}.

The all-loop asymptotic Bethe equations for the length $2L$ spin chain can be obtained from \cite{Gromov:2008qe} by restricting to the $SU(2)\times SU(2)$ flavor sector
\begin{equation}\label{BAEsu2}\begin{split}
\left[\frac{x(u_j+i/2)}{x(u_j-i/2)}\right]^L=& \prod_{k=1,k\neq
j}^{M_u}\frac{u_j-u_k+i}{u_j-u_k-i}\,e^{i\theta(u_j,u_k)}\,\prod_{k=1}^{M_v}e^{i\theta(u_j,v_k)}\, ,\\
\left[\frac{x(v_j+i/2)}{x(v_j-i/2)}\right]^L=& \prod_{k=1,k\neq
j}^{M_v}\frac{v_j-v_k+i}{v_j-v_k-i}\,e^{i\theta(v_j,v_k)}\,\prod_{k=1}^{M_u}e^{i\theta(v_j,u_k)}\, ,
\end{split}\end{equation}
and must be supplemented by  the momentum constraint
\begin{equation}
\prod_{j=1}^{M_u}\frac{x(u_j+i/2)}{x(u_j-i/2)}\,\prod_{j=1}^{M_v}\frac{x(v_j+i/2)}{x(v_j-i/2)}=1\,
.
\end{equation}
We have called $u_i$ and $v_i$ the Bethe roots of each $SU(2)$
factor. We also denoted by $M_u$ and $M_v$ the number of $u$ and
$v$ roots. Moreover, we introduced the function
\begin{equation}
x(w)=\frac{w}{2}\left(1+\sqrt{1-4\frac{h^2(\lambda)}{w^2}}\right)
\, ,
\end{equation}
where $w=u,v$. As we see from (\ref{BAEsu2}), the dressing factor
$e^{i\theta}$ introduces extra self-interactions for the roots and
also couples the two $SU(2)$ spin chains at higher loops.
Following \cite{Gromov:2008qe} we define the dressing phase as in
\cite{Beisert:2005wv,Beisert:2006ez}
\begin{equation}\label{dressingphase}
\theta(w_i,w_j)=\sum_{r=2}^{\infty}\sum_{s=r+1}^{\infty}\beta_{r,s}(\lambda)[q_r(w_i)q_s(w_j)-q_s(w_i)q_r(w_j)]\,
,
\end{equation}
where the coefficient functions
$\beta_{r,s}(\lambda)$ can be expanded in $\lambda$ as
\begin{equation}\label{pertdressing}
\beta_{r,s}(\lambda)=\sum_{k=s-1}^{\infty}\beta_{r,s}^{(2k)}\lambda^{2k}
\end{equation}
and the quantities
\begin{equation}
q_r(w)=\frac{1}{r-1}\left(\frac{i}{x(w+i/2)^{r-1}}-\frac{i}{x(w-i/2)^{r-1}}\right)
\end{equation}
are related to the eigenvalues of the conserved charges of the
theory, whose existence is ensured by integrability. In
particular, the second charge is the hamiltonian of the integrable
model and is identified with the dilatation operator of the gauge
theory. The energy eigenvalues are given by
\begin{equation}
E=h^2(\lambda)\left(\sum_{j=1}^{M_u}q_2(u_j)+\sum_{j=1}^{M_v}q_2(v_j)\right)\,
.
\end{equation}
Let's observe that, for two-impurity states, the Bethe Ansatz is simplified.
The momentum constraint requires
\begin{equation}
 w_1=-w_2\equiv w\, ,
\end{equation}
while the Bethe equations reduce to
\begin{equation}
 \left(\frac{w+i/2}{w-i/2}\right)^L=\,\left(\frac{1+\sqrt{1-\frac{4h^2(\lambda)}{(w-i/2)^2}}}{1+\sqrt{1-\frac{4h^2(\lambda)}{(w+i/2)^2}}}\right)^L\,e^{i\theta(w,-w)}
\end{equation}
for $M_u=M_v=1$, and
\begin{equation}
 \left(\frac{w+i/2}{w-i/2}\right)^{L-1}=\,\left(\frac{1+\sqrt{1-\frac{4h^2(\lambda)}{(w-i/2)^2}}}{1+\sqrt{1-\frac{4h^2(\lambda)}{(w+i/2)^2}}}\right)^L\,e^{i\theta(w,-w)}
\end{equation}
for $M_u=2$, $M_v=0$ or $M_u=0$, $M_v=2$. For any fixed $L$, these equations can be solved order by order in perturbation theory. This is the basic tool for the computation of anomalous
dimensions of long operators.

As can be seen from the Bethe equations, the anomalous dimensions
of operators will depend, in general, on the coefficients of the
function $h(\lambda)$ and of the dressing phase. It seems that
such coefficients cannot be determined using the integrability of
the theory only. However, some general considerations on their
values can be made, as we now explain. In the case of
$\mathcal{N}=4$ SYM theory it was shown \cite{Beisert:2006ez}
that the coefficients of the dressing phase should have a well
defined degree of transcendentality in order to preserve the
Kotikov-Lipatov transcendentality principle on the scaling
function of the theory \cite{Kotikov:2002ab}. Such principle can
be generalized to the ABJM case: the scaling function of ABJM
theory is prescribed to be \cite{Gromov:2008qe}
\begin{equation}
f_{\text{ABJM}}(\lambda)=\frac{1}{2}\, f_{\text{SYM}}(g)|_{g\rightarrow
h(\lambda)}\, ,
\end{equation}
where $g=\frac{\sqrt{\lambda}}{4\pi}$ in $\mathcal{N}=4$ SYM.
Taking the weak coupling expansion for $f_{\text{SYM}}(g)$ obtained in
\cite{Beisert:2006ez} and inserting the expansion of $h(\lambda)$,
we thus get, up to eight loops,
\begin{equation}\label{cusp}\begin{split}
f_{\text{ABJM}}(\lambda)=\,&4\,\lambda^2-\left(\frac{4}{3}\pi^2-4h_4\right)\,\lambda^4+\left(\frac{44}{45}\pi^4-\frac{8}{3}h_4\pi^2+4h_6\right)\,\lambda^6\\
&-4\,\left(\frac{73}{315}\pi^6-8\zeta(3)^2-\frac{11}{15}h_4\pi^4+\frac{1}{3}h_4^2\pi^2+\frac{2}{3}h_6\pi^2-h_8+4\beta_{2,3}^{(6)}\zeta(3)\right)\,\lambda^8\\
&+\mathcal{O}(\lambda^{10})\, .
\end{split}\end{equation}
In deriving (\ref{cusp}) we have used the fact that
$\beta_{r,s}=0$ for $r+s$ even and that, in our case, the
coefficients $\beta_{r,s}(\lambda)$ should contribute starting at
order $\mathcal{O}(\lambda^{2(r+s-2)})$ rather than
$\mathcal{O}(\lambda^{2(s-1)})$, as expected from
(\ref{pertdressing}). See \cite{Beisert:2005wv,Beisert:2006ez} for
an explanation. The latter property will be verified, for the $\beta_{2,3}$ 
coefficient, in Section \ref{dilop}. The transcendentality principle, extended to ABJM
theory, states
that:\vspace{0.2cm}\\
\emph{Assigning degree of transcendentality $k$ to constants
$\pi^k$ and $\zeta(k)$, the $\ell$-loop contribution to the
scaling function $f_{ABJM}$ has uniform degree of
transcendentality $\ell-2$}.\vspace{0.2cm}

Recalling that $h_4=-4\zeta(2)$, we see that this principle is
satisfied up to four loops. Moreover, from the six-loop
contribution, we immediately see that $h_6$ should have degree
four. Starting from eight loops also the unknown dressing phase
starts contributing to the scaling function. One very natural
way\footnote{We're neglecting the possibility that $h_8$ and
$\beta_{2,3}^{(6)}$ can have higher (or lower) degrees of
transcendentality which cancel out between them.} to preserve the
transcendentality principle is to conjecture that:
\begin{itemize}
    \item $h_{2k}$ has degree of transcendentality $2k-2$,
    \item $\beta_{r,s}^{(2k)}$ has degree of transcendentality
    $2k+2-r-s$.
\end{itemize}
In particular, we notice that the transcendentality principle
implies that the leading order coefficient of the dressing phase
should be of the form
\begin{equation}
\beta_{2,3}^{(6)}\,=\,a\,\pi^3+b\,\zeta(3)\, ,
\end{equation}
where $a$ and $b$ are some rational numbers. The main goal of this paper is to
compute such constants.

\section{Construction of the Dilatation Operator}\label{dilop}
The dilatation operator is the generator of scaling
transformations and measures the scaling dimensions of composite
operators. The perturbative expansion of the dilatation operator
in the $SU(2)\times SU(2)$ sector is
\begin{equation}\label{Dexp}
\mathcal{D}(\lambda)=L+\sum_{k=1}^{\infty}\,\mathcal{D}_{2k}\,\lambda^{2k}\,
,
\end{equation}
where the 0-loop term $L$ is proportional to the identity and
yields the classical dimension of an operator made up of $2L$
chiral superfields, while the second term is the quantum
contribution and yields its anomalous dimension. The latter can be
extracted from the perturbative renormalization of the composite
operators $\mathcal{O}_a$
\begin{equation}
\mathcal{O}_{a}^{\text{ren}}=\mathcal{Z}_{a}{}^b\mathcal{O}_{b}^{\text{bare}}\,
,\;\;\;\;\;\;\mathcal{Z}=\mathds{1}+\lambda^2\mathcal{Z}_2+\lambda^4\mathcal{Z}_4+\cdots\,
.
\end{equation}
The renormalization factor $\mathcal{Z}$ is introduced in order to
remove UV divergences from correlation functions of operators and
can be computed in perturbation theory by means of standard
methods. Using dimensional regularization in spacetime dimension
$D=3-2\varepsilon$, quantum divergences show up as inverse powers
of $\varepsilon$ in the limit  $\varepsilon\to0$. By introducing
the 't Hooft mass $\mu$ and the dimensionful combination
$\lambda\mu^{2\varepsilon}$ the dilatation operator is then
extracted from $\mathcal{Z}$ as
\begin{equation}\label{DZ}
\mathcal{D}= L
+\mu\frac{\d}{\d\mu}\ln\mathcal{Z}(\lambda\mu^{2\varepsilon},\varepsilon)
=L+\lim_{\varepsilon\rightarrow0}\left[2\varepsilon\lambda
\frac{\d}{\d\lambda}\ln\mathcal{Z}(\lambda,\varepsilon)\right]\, .
\end{equation}
This definition effectively extracts the $1/\varepsilon$
pole of $\ln\mathcal{Z}$; the higher order
poles must be absent. A full-fledged quantum field theory
computation of the dilatation operator can be an extremely
difficult task, especially for high loop orders in perturbation
theory. Fortunately, things get simplified if we make use of the
integrability results.

Here we discuss a general procedure to construct the asymptotic
dilatation operator of ABJM theory. It is similar to the procedure
used in the case of $\mathcal{N}=4$ SYM and described in
\cite{Beisert:2003tq, Beisert:2004hm, Beisert:2004ry, Beisert:2005wv, Fiamberti:2010ra}.
The idea is the following: starting from an exactly integrable
hamiltonian (which is next-to-nearest neighbor for ABJM theory),
we deform it with local interactions of range linearly increasing
with the perturbative order, obtaining a long-ranged spin chain.
At $\ell$ loops the maximum range is $\ell+1$. When the range of
an interaction exceeds the length $2L$ of the spin chain,
\emph{wrapping interactions} typically appear and the Bethe Ansatz
breaks down. So, we work with \emph{asymptotic states},
\emph{i.e.} states of length $2L>\ell$.

The two-loop dilatation operator \cite{Minahan:2008hf, Bak:2008cp} is the sum
of two XXX$_{\frac{1}{2}}$ Heisenberg hamiltonians, one living on
odd and the other on even sites: this hamiltonian is deformed with
interactions of longer range at higher perturbative orders.

The dilatation operator is conveniently written in terms of the basis of
\emph{permutation structures}
\begin{equation}
 \{a_1,\ldots,a_n\}=\sum_{j=1}^L\,P_{2j+a_1,2j+a_1+2}\cdots
 P_{2j+a_n,2j+a_n+2}\, ,
\end{equation}
where $P_{i,j}$ are permutation operators. Permutation structures
represent local interactions of spins summed over all different
positions on the spin chain: their properties are collected in
Appendix \ref{PermutationStructures}. Here, we just need to define  the number
\begin{equation}
\mathcal{R}=\max(a_1,\ldots,a_n)-\min(a_1,\ldots,a_n)+3
\end{equation}
which is the \emph{range} of the interaction.
\begin{table}
\centering
\begin{tabular}{|c|c|}
  \hline
  $\mathcal{R}$ & \textbf{Basis of permutation structures} \\
  \hline
  \hline
  1 & \{\,\} \\
  \hline
  2 & - \\
  \hline
  3 & \{0\} \\
  \hline
  4 & \{0,1\} \\
  \hline
  5 & \{0,2\}, \{2,0\} \\
    & (\{0,1,2\}, \{2,1,0\}) \\
  \hline
  6 & \{0,3\} \\
    & \{0,1,3\}, \{0,3,1\} \\
    & \{0,2,3\}, \{2,0,3\}\\
    &  (\{0,1,2,3\}, \{2,1,0,3\}, \{0,3,2,1\}, \{2,3,0,1\})  \\
  \hline
  7 &  \{0,4\}\\
    &  \{0,2,4\}, \{4,2,0\}, \{0,4,2\}, \{2,0,4\} \\
    & (\{0,1,4\}, \{0,3,4\})\\
    & (\{2,0,4,2\})\\
    & (\{0,1,2,4\}, \{2,1,0,4\}, \{4,1,2,0\}, \{4,1,0,2\})\\
    & (\{0,1,3,4\}, \{0,3,1,4\}) \\
    & (\{0,2,3,4\}, \{2,0,3,4\}, \{0,4,3,2\}, \{4,3,2,0\}) \\
  \hline
  8 & \{0,5\}, \{0,1,5\}, \{0,4,5\}\\
    & \{0,2,5\}, \{2,0,5\}, \{0,3,5\}, \{0,5,3\}\\
    & \{0,1,3,5\}, \{0,3,1,5\}, \{0,1,5,3\}, \{5,3,1,0\} \\
    & \{0,2,3,5\}, \{2,0,3,5\}, \{0,2,5,3\}, \{2,0,5,3\} \\
    & \{0,2,4,5\}, \{2,0,4,5\}, \{0,4,2,5\}, \{4,2,0,5\} \\
  \hline
  9 & \{0,6\} \\
    & \{0,2,6\}, \{2,0,6\}, \{0,4,6\}, \{0,6,4\} \\
    & \{0,3,6\} \\
    & \{0,2,4,6\}, \{0,2,6,4\}, \{0,4,2,6\}, \{2,0,4,6\} \\
    & \{2,0,6,4\}, \{0,6,4,2\}, \{4,2,0,6\}, \{6,4,2,0\} \\
  \hline
\end{tabular}
\vspace{0.5cm} \caption{Permutation structures
needed  up to eight loops,  grouped according to their range $\mathcal{R}$. Permutation
structures of odd (even) range $\mathcal{R}$ appear at $\mathcal{R}-1$ ($\mathcal{R}$) loops,
apart from the ones in parenthesis, which appear at the next loop
order. Permutation structures contributing at ten loops or beyond
are not written. Note that, in order to find the complete basis,
one has to add $\{\ldots,a+1,\ldots\}$ to each permutation
structure $\{\ldots,a,\ldots\}$ listed here.} \label{permbasis}
\end{table}

In Table \ref{permbasis} we list all the independent permutation
structures that appear in the dilatation operator up to eight
loops. In order to read a basis for the dilatation operator at
$\ell$ loops, one has to include all the permutation structures up
to range $\mathcal{R}=\ell +1$, neglecting the ones in parenthesis, because
they appear at the next loop order.

In order to construct the $\ell$-loop dilatation operator, we
first write the most general combination of permutation structures
up to range $\ell+1$. Then, we fix a large part of the unknown coefficients as follows:
\begin{enumerate}
  \item Impose hermiticity and parity invariance:
\begin{displaymath}
 \mathcal{D}_\ell^\dagger=\mathcal{D}_\ell\,
 ,\;\;\;\;\;\;\;\;\;\;\;\;\;\;\mathcal{P}\mathcal{D}_\ell\mathcal{P}^{-1}=\mathcal{D}_\ell\,
 ;
\end{displaymath}
 \item Require that the vacuum state is protected (\emph{i.e.}, it has zero energy):
\begin{displaymath}
 \mathcal{D}_\ell\,|0\,\rangle=0\, ;
\end{displaymath}
 \item Impose the magnon dispersion relation on one-impurity states on odd and even
sites (let $E_\ell$ be the $\ell$-loop coefficient of
(\ref{dispersion}) in the small $\lambda$ expansion):
\begin{displaymath}
 \mathcal{D}_\ell\,|p\rangle_{o,e}=E_\ell(p)\,|p\rangle_{o,e}\, ;
\end{displaymath}
 \item Use the asymptotic Bethe ansatz on two-impurity states in order to fix the remaining
parameters.
\end{enumerate}

Point 4 works in the following manner:
\begin{itemize}
 \item Fix a length of the spin chain, with $L>\ell/2$, so that wrapping
interactions are absent;
 \item Find a basis of two-impurity operators of length $2L$. Note that there are two types
 of such bases, as the impurities can be both on even (or odd) sites or one on even and the other on odd sites\footnote{For the same reason, we
 have seen in section \ref{BA} that there are two types of Bethe equations for two-impurity states.};
 \item Explicitly diagonalize the matrix representation of $\mathcal{D}(\lambda)=\sum_{k=1}^{\ell/2}\,\mathcal{D}_{2k}\,\lambda^{2k}$
 on the basis of length $2L$ states;
 \item Compute the eigenvalues from the asymptotic Bethe equations (\ref{BAEsu2}) and
compare them with those found in the previous step, containing the
unknown parameters.
\end{itemize}

The explicit diagonalization of the dilatation operator and the
solution of the Behte equations can be performed with the help of
\texttt{Mathematica}.

At the end of the procedure, some parameters may remain unfixed.
Some of them are related to \emph{similarity transformations} of
the dilatation operator:
\begin{equation}
\mathcal{D}'=e^{-i\chi}\mathcal{D}e^{i\chi}\, .
\end{equation}
They don't appear in the spectrum but only affect the eigenvectors.
Therefore, they can be regarded as unphysical and their values
depend on the renormalization scheme.

Another way to fix the unknown parameters is through conservation of
the first higher charge $\mathcal{Q}_3$, if we assume perturbative
integrability. This requires the construction of $\mathcal{Q}_3$
up to order $\ell$ with a procedure similar to that described for
the dilatation operator, with the difference that
 $\mathcal{Q}_3^{(\ell)}$ has odd parity and is anti-hermitian. Since
$\mathcal{Q}_3$ has higher maximal range, $\ell+2$, at high loops
it can be quite complicated to compute.
Some other parameters may depend on the
dressing phase and cannot be fixed through the Bethe Ansatz.

We now apply the procedure described above to constrain the form of the dilatation operator up to eigth loops. We immediately get
\begin{equation}\label{dilatation2}
 \mathcal{D}_2=2\,\{\,\}-\{0\}-\{1\}
\end{equation}
for the two-loop dilatation operator and
\begin{equation}\label{dilatation4}
\mathcal{D}_4=2\,(-4+h_4)\,\{\,\}+(6-h_4)\,(\{0\}+\{1\})-(\{0,2\}+\{2,0\}+\{1,3\}+\{3,1\})
\end{equation}
for the four-loop dilatation operator. We see that
(\ref{dilatation2}) reproduces the result of 
\cite{Minahan:2008hf, Bak:2008cp}, while (\ref{dilatation4}) reproduces the
results\footnote{The computation of \cite{Minahan:2009aq} followed
a procedure similar to ours, up to point 3. This fixes all but one
coefficient in the dilatation operator, namely the coefficient of
$\{0,1\}+\{1,2\}$. We simply use point 4 of our procedure on
length six two-impurity states to prove that it vanishes, without
computing any Feynman diagram.} of \cite{Minahan:2009aq} and
\cite{Bak:2009mq}.

The six-loop dilatation operator was computed in \cite{Bak:2009tq}
through Feynman diagrammatics. We apply our technique to rederive
it. 
Assuming that the dressing phase is absent at this loop order,
following \cite{Beisert:2005wv,Beisert:2006ez}, we can completely fix the dilatation operator
with our procedure.
If we don't assume that the dressing phase is absent, a single input
from the results of \cite{Bak:2009tq} is sufficient to determine the
full dilatation operator. We follow this latter approach.
First of all we write the
six-loop dilatation operator in the basis of permutation
structures. After having imposed hermiticity and parity
invariance, it is given by
\begin{equation}\begin{split}
\mathcal{D}_6=&a\,\{\,\}+\mathcal{D}_{6,even}+\mathcal{D}_{6,odd}+\mathcal{D}_{6,mixed}\, ,\\
\end{split}\end{equation}
where
\begin{equation}\begin{split}
\mathcal{D}_{6,even}=\,&\,
b\,\{0\}+d\,(\{0,2\}+\{2,0\})+m\,\{0,4\}\\
&+l\,(\{0,2,4\}+\{4,2,0\})+\tilde{l}\,(\{2,0,4\}+\{0,4,2\})\\
&+i\,\epsilon_1\,(\{2,0,4\}-\{0,4,2\})\, ,
\end{split}\end{equation}

\begin{equation}
\mathcal{D}_{6,odd}=\mathcal{D}_{6,even}(\{\ldots,a,\ldots\}\leftrightarrow\{\ldots,a+1,\ldots\})\,
,\phantom{XXXX}
\end{equation}

\begin{equation}\begin{split}
\mathcal{D}_{6,mixed}=\,&\,
c\,(\{0,1\}+\{1,2\})+e\,(\{0,3\}+\{1,4\})\\
&+f\,(\{0,1,2\}+\{2,1,0\}+\{1,2,3\}+\{3,2,1\})\\
&+g\,(\{0,1,3\}+\{0,3,1\}+\{1,3,4\}+\{3,1,4\}\\
&\phantom{XX}+\{0,2,3\}+\{2,0,3\}+\{1,4,2\}+\{1,2,4\})\\
&+i\,\epsilon_2\,(\{0,1,3\}-\{0,3,1\}-\{1,3,4\}+\{3,1,4\}\\
&\phantom{XX}-\{0,2,3\}+\{2,0,3\}-\{1,4,2\}+\{1,2,4\})\, ,\\
\end{split}\end{equation}
where all parameters are real. Zero-energy of the vacuum gives
the equation
\begin{equation}
 a+2b+4d+4l+4\tilde{l}+2m+2c+2e+4f+8g=0\, .
\end{equation}
We then compute the action of $\mathcal{D}_{6}$ on $|p\rangle_e$ and $|p\rangle_o$
and compare the result with the six loop coefficient of the
weak-coupling expansion of (\ref{dispersion}) obtaining the
equations

\begin{equation}\begin{split}
l=&\,-2\, ,\\
b=&\,-48-2c-2e-2f-4g+12h_4-h_6+2\tilde{l}-2m\, ,\\
d=&\,12-f-2g-2h_4-2\tilde{l}\, .
\end{split}\end{equation}
Now we consider two-impurity states. In order to avoid wrapping, we have to start
from $L=4$. For $L=4$ (length 8 states), comparing the eigenvalues
of the dilatation operator with the solutions of the Bethe
equations we get
\begin{equation}\label{mbeta}
\tilde{l}=0\, ,\;\;\;\;\;\;\;2m=4-3\,\beta_{2,3}^{(4)}\,
,\;\;\;\;\;\;\;f=0\, ,\;\;\;\;\;\;\;g=0\, ,\;\;\;\;\;\;\;c+e=0\, .
\end{equation}
The results of \cite{Bak:2009tq} (obtained with an independent
approach) imply, in our notations, that $m=2$. From (\ref{mbeta})
we thus see that $\beta_{2,3}^{(4)}$ must vanish. For $L=5$
(length 10 states) we get
\begin{equation}
c=0\, ,\;\;\;\;\;\;\;\;e=0\, .
\end{equation}
We therefore have explicitly shown that, up to irrelevant parameters,
$\mathcal{D}_{6,mixed}=0$. It means that the two types of magnons
don't interact at six loops, \emph{i.e.} the two $SU(2)$ factors
are decoupled, as claimed in \cite{Gromov:2008qe}.

The six-loop dilatation operator is then given by
\begin{equation}\begin{split}
\mathcal{D}_6=\,&\,2\,(30-8h_4+h_6)\,\{\,\}+(-52+12h_4-h_6)\,(\{0\}
+\{1\})\\
&+2\,(\{0,4\}+\{1,5\})+(12-2h_4)\,(\{0,2\}+\{2,0\}+\{1,3\}+\{3,1\})\\
&-2\,(\{0,2,4\}+\{4,2,0\}+\{1,3,5\}+\{5,3,1\})\\
&+i\,\epsilon_1\,(\{2,0,4\}-\{0,4,2\}-\{1,5,3\}+\{3,1,5\})\\
&+i\,\epsilon_2\,(\{0,1,3\}-\{0,3,1\}-\{1,3,4\}+\{3,1,4\}\\
&\phantom{XX}-\{0,2,3\}+\{2,0,3\}-\{1,4,2\}+\{1,2,4\})\, .\\
  \end{split}\end{equation}
Let's observe that, setting the unphysical coefficients
$\epsilon_1$ and $\epsilon_2$ to zero, $\mathcal{D}_6$ coincides
with the operator presented in \cite{Bak:2009tq}. The
coefficient $\beta_{2,3}^{(4)}$, which can be a priori present
according to (\ref{pertdressing}), actually vanishes. Therefore,
the first contribution to the dressing phase should be searched at
eight loops. We are thus led to consider the eight-loop dilatation
operator: since it isn't known at present, we now use our
procedure to constrain it as much as possible.

The permutation structures contributing to $\mathcal{D}_8$ are
listed in Table \ref{permbasis}. After imposing hermiticity and
parity invariance, the eight-loop dilatation operator can be
written (with real coefficients) as
\begin{equation}\begin{split}
\mathcal{D}_8=&a\,\{\,\}+\mathcal{D}_{8,even}+\mathcal{D}_{8,odd}+\mathcal{D}_{8,mixed}\, ,
\end{split}\end{equation}
where
\begin{equation}\begin{split}
\mathcal{D}_{8,even}=\,&
b\,\{0\}+c\,(\{0,2\}+\{2,0\})+d\,\{0,4\}\\
&+e_1\,(\{0,2,4\}+\{4,2,0\})+e_2\,(\{0,4,2\}+\{2,0,4\})\\
&+i\,\epsilon_3\,(\{0,4,2\}-\{2,0,4\})+f\,\{0,6\}\\
&+g\,(\{0,2,6\}+\{2,0,6\}+\{0,4,6\}+\{0,6,4\})\\
&+i \epsilon_1\,(\{0,2,6\}-\{2,0,6\}-\{0,4,6\}+\{0,6,4\})\\
&+l_1\,(\{0,2,4,6\}+\{6,4,2,0\})\\
&+l_2\,(\{0,2,6,4\}+\{2,0,4,6\}+\{0,6,4,2\}+\{4,2,0,6\})\\
&+i\,\epsilon_2\,(\{0,2,6,4\}-\{2,0,4,6\}+\{0,6,4,2\}-\{4,2,0,6\})\\
&+l_3\,(\{0,4,2,6\}+\{2,0,6,4\})+m\,\{2,0,4,2\}\, ,
\end{split}\end{equation}

\begin{equation}
 \begin{split}
  \mathcal{D}_{8,mixed}=\,&
\alpha_1\,\{0,1\}+\alpha_2\,(\{0,1,2\}+\{2,1,0\})+\alpha_3\,\{0,3\}\\
&+\alpha_4\,(\{0,1,3\}+\{0,3,1\}+\{0,2,3\}+\{2,0,3\})\\
&+i\,\epsilon_{m1}\,(\{0,1,3\}-\{0,3,1\}-\{0,2,3\}+\{2,0,3\})\\
&+\alpha_5\,(\{0,1,2,3\}+\{2,3,0,1\})+\alpha_{6}\,(\{2,1,0,3\}+\{0,3,2,1\})\\
&+i\,\epsilon_{m2}\,(\{2,1,0,3\}-\{0,3,2,1\})+\alpha_7\,(\{0,1,4\}+\{0,3,4\})\\
&+\alpha_{8}\,(\{0,1,2,4\}+\{4,1,2,0\}+\{0,2,3,4\}+\{4,3,2,0\})\\
&+i\,\epsilon_{m3}\,(\{0,1,2,4\}-\{4,1,2,0\}-\{0,2,3,4\}+\{4,3,2,0\})\\
&+\alpha_{9}\,(\{2,1,0,4\}+\{4,1,0,2\}+\{2,0,3,4\}+\{0,4,3,2\})\\
&+i\,\epsilon_{m4}\,(\{2,1,0,4\}-\{4,1,0,2\}+\{2,0,3,4\}-\{0,4,3,2\})\\
&+\alpha_{10}\,(\{0,1,3,4\}+\{0,3,1,4\})+\alpha_{11}\,\{0,5\}\\
&+\alpha_{12}\,(\{0,1,5\}+\{0,4,5\})\\
&+\alpha_{13}\,(\{0,2,5\}+\{2,0,5\}+\{0,3,5\}+\{0,5,3\})\\
&+i\,\epsilon_{m5}\,(\{0,2,5\}-\{2,0,5\}-\{0,3,5\}+\{0,5,3\})\\
&+\alpha_{14}\,(\{0,1,3,5\}+\{4,2,0,5\}+\{5,3,1,0\}+\{0,2,4,5\})\\
&+i\,\epsilon_{m6}\,(\{0,1,3,5\}+\{4,2,0,5\}-\{5,3,1,0\}-\{0,2,4,5\})\\
&+\alpha_{15}\,(\{0,3,1,5\}+\{0,1,5,3\}+\{2,0,4,5\}+\{0,4,2,5\})\\
&+i\,\epsilon_{m7}\,(\{0,3,1,5\}-\{0,1,5,3\}+\{2,0,4,5\}-\{0,4,2,5\})\\
&+\alpha_{16}\,(\{0,2,3,5\}+\{2,0,5,3\})+\alpha_{17}\,(\{2,0,3,5\}+\{0,2,5,3\})\\
&+i\,\epsilon_{m8}\,(\{2,0,3,5\}-\{0,2,5,3\})\\
&+\alpha_{18}\,\{0,3,6\}+(\{\ldots,a\leftrightarrow
a+1,\ldots\})\, .
 \end{split}
\end{equation}
Requiring that the vacuum state is protected we have one condition
on the coefficients. If we then impose the eight-loop dispersion
relation on one-magnon states we obtain four more conditions,
which in particular fix the coefficient $l_1=-5$. In order to fix
other parameters we consider two-impurity states: for sufficiently
long states in order to avoid wrapping effects, with the help of \texttt{Mathematica}, we
explicitly diagonalize the dilatation operator and compare its
eigenvalues with those obtained from the perturbative solutions of
the asymptotic Bethe equations. For $L=5,6,7$ (length 10, 12, 14
states) we find 12 more conditions, while higher length states
give no more equations. So the point 4 of the procedure
described above isn't sufficient to completely fix the dilatation
operator. Nevertheless, we stress that an important result already
follows from the equations we could derive from point 4: a non
vanishing\footnote{We will see in Section \ref{supergraphcompute}
that this is, indeed, the case.} coefficient $\beta_{2,3}^{(6)}$
for the dressing phase would imply that the $\alpha_i$
coefficients, appearing in $\mathcal{D}_{8,mixed}$, cannot be all
zero. This proves that $\mathcal{D}_{8,mixed}\neq 0$. In
principle, the undetermined coefficients could be fixed imposing
the commutation of $\mathcal{D}$ with the third charge
$\mathcal{Q}_3$: this is laborious and we won't do it, since it
isn't necessary for the computations of the dressing phase, as we
will see in the next chapter. We quote here the part of the
dilatation operator which acts on even sites only and exchanges
the fields inside the operator in a maximal way:
\begin{equation}\label{dilatationeven}
 \begin{split}
  \mathcal{D}_{8,mr}=
             &+m\,\{2,0,4,2\}\\
             \vspace{0.4cm}
             &+l_3\,(\{0,4,2,6\}+\{2,0,6,4\})\\
             &+\Big(-1+\frac{1}{2}m+i\,\epsilon_{2a}+\frac{1}{4}\beta\Big)\,(\{0,2,6,4\}+\{0,6,4,2\})\\
             &+\Big(-1+\frac{1}{2}m-i\,\epsilon_{2a}+\frac{1}{4}\beta\Big)\,(\{2,0,4,6\}+\{4,2,0,6\})\\
             \vspace{0.4cm}
             &-5\,(\{0,2,4,6\}+\{6,4,2,0\})
 \end{split}
\end{equation}
where the coefficient $\epsilon_{2a}$ corresponds to similarity
transformations and depends on the particular renormalization 
scheme used. The dressing phase enters also
$\mathcal{D}_{8,mr}$ through the parameter
$\beta=\beta_{2,3}^{(6)}$. The form of (\ref{dilatationeven}) is
the same of the $\mathcal{N}=4$ SYM counterpart in the four-loop
dilatation operator \cite{Beisert:2007hz}. It is interesting to
note that the coefficients $\alpha_i$ and the coefficients of
$h(\lambda)$ don't appear in (\ref{dilatationeven}). However, the
rest of the dilatation operator contains also these parameters: in
general, the ``mixed'' hamiltonian acts coupling the two types of
magnons; this is a completely new feature of ABJM theory which
appears at eight loops, making this theory substantially different
from the $\mathcal{N}=4$ SYM theory at four loops.

\section{Computation of the Dressing Phase}\label{supergraphcompute}
In this Section we proceed to the computation of the leading-order
coefficient of the dressing phase, \emph{i.e.} the parameter
$\beta_{2,3}^{(6)}$ which appears in the eight-loop dilatation
operator. We determine it by direct field-theory calculations, in
a similar fashion to the $\mathcal{N}=4$ SYM case
\cite{Beisert:2007hz}. Here, we use $\mathcal{N}=2$ superspace
methods for the evaluation of Feynman diagrams. We refer to
Appendix \ref{superspace} for a very brief review of such methods.

\begin{figure}
\centering
\parbox{3cm}{\centering\includegraphics[width=2.5cm, height=2cm]{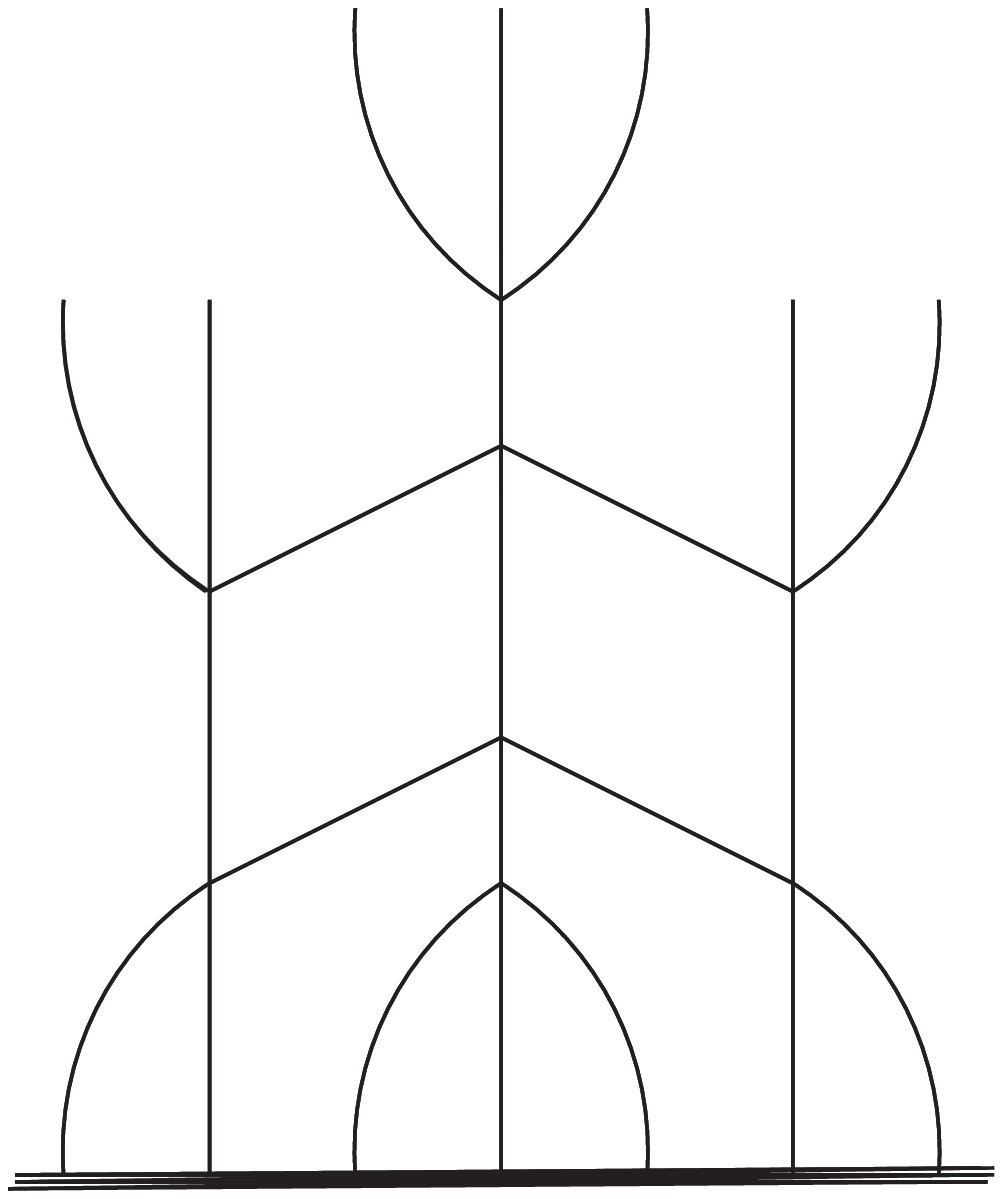}\\ $\mathcal{S}_a$}
\parbox{3cm}{\centering\includegraphics[width=2.5cm, height=2cm]{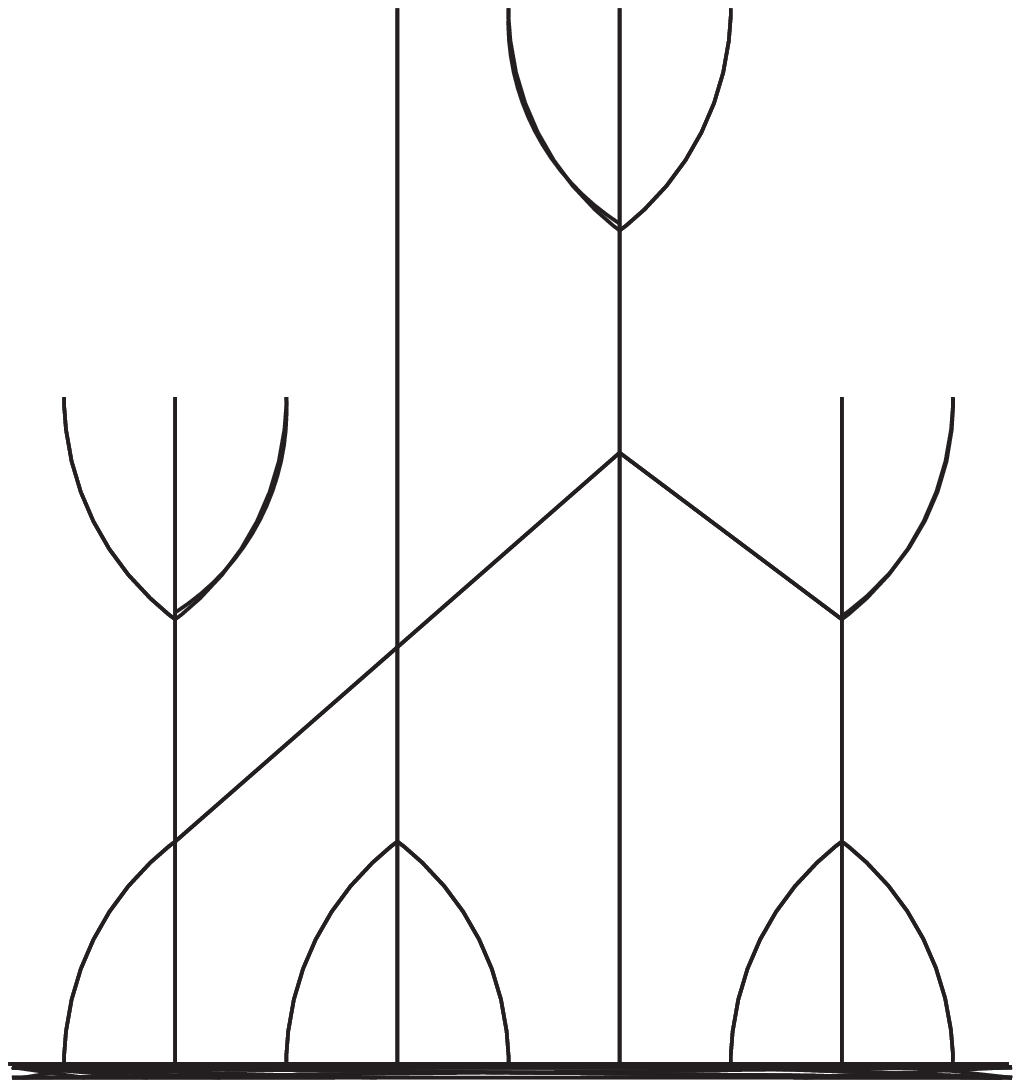}\\ $\mathcal{S}_b$}
\parbox{3cm}{\centering\includegraphics[width=2.5cm, height=2cm]{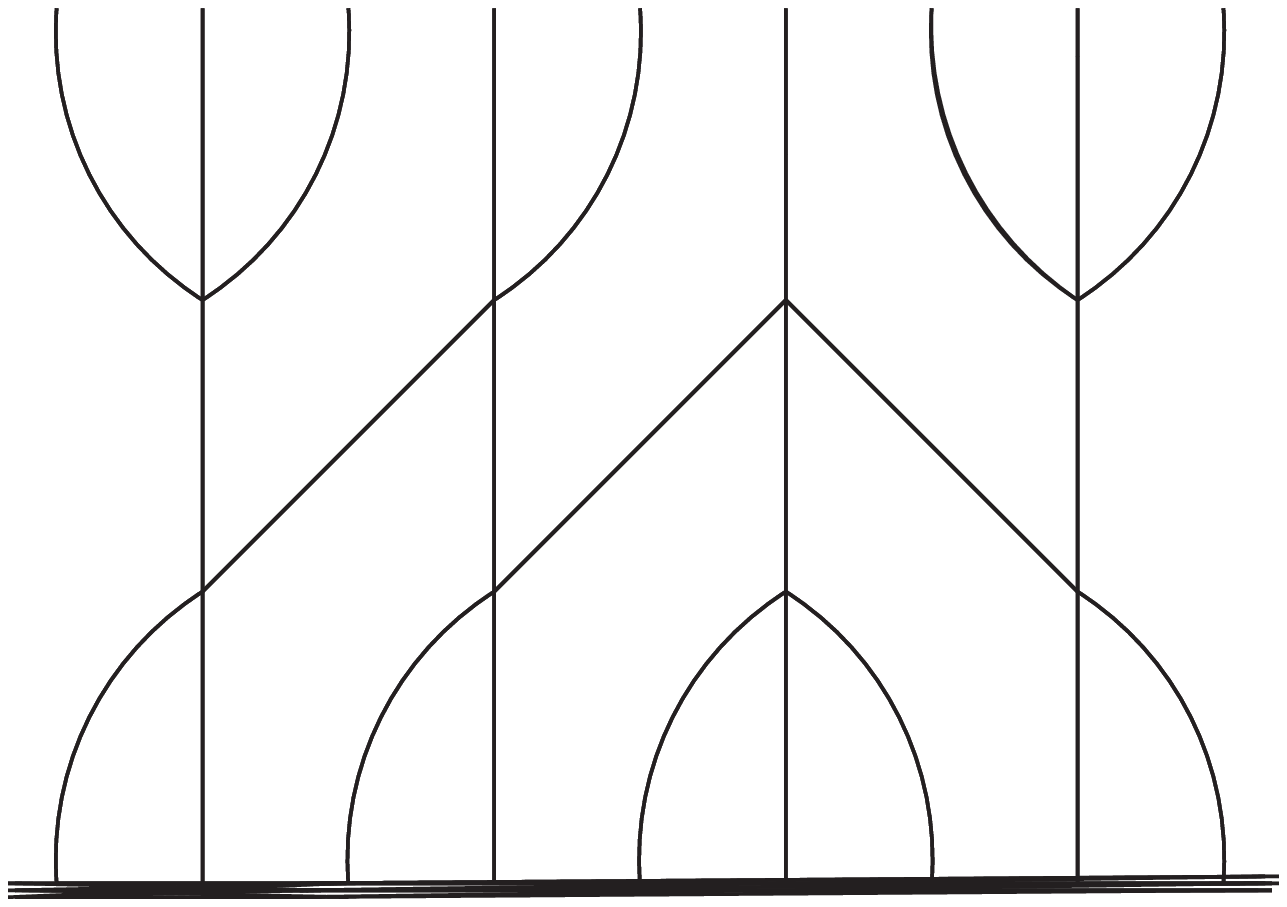}\\ $\mathcal{S}_c$}
\parbox{3cm}{\centering\includegraphics[width=2.5cm, height=2cm]{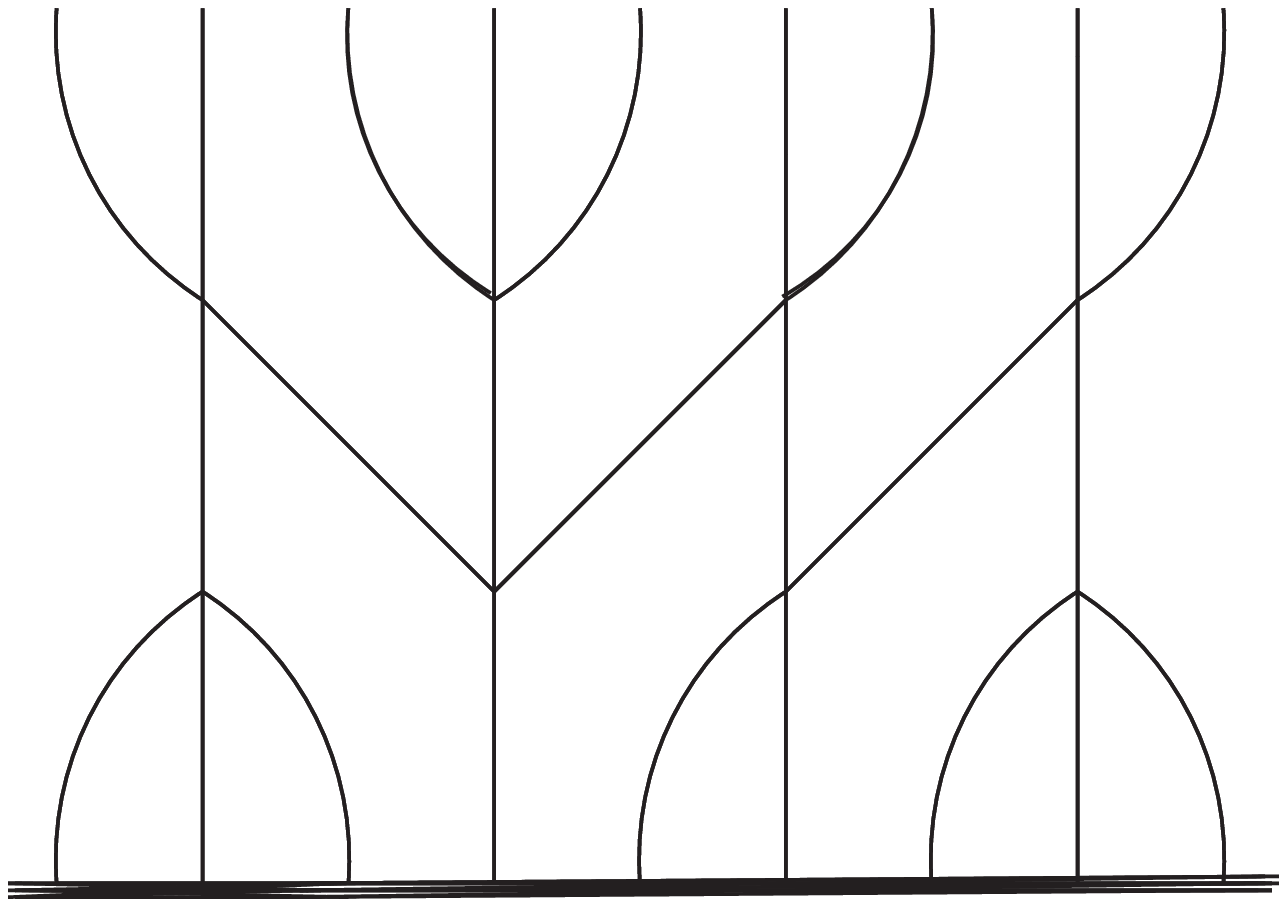}\\ $\mathcal{S}_d$}
\parbox{3cm}{\centering\includegraphics[width=2.5cm, height=2cm]{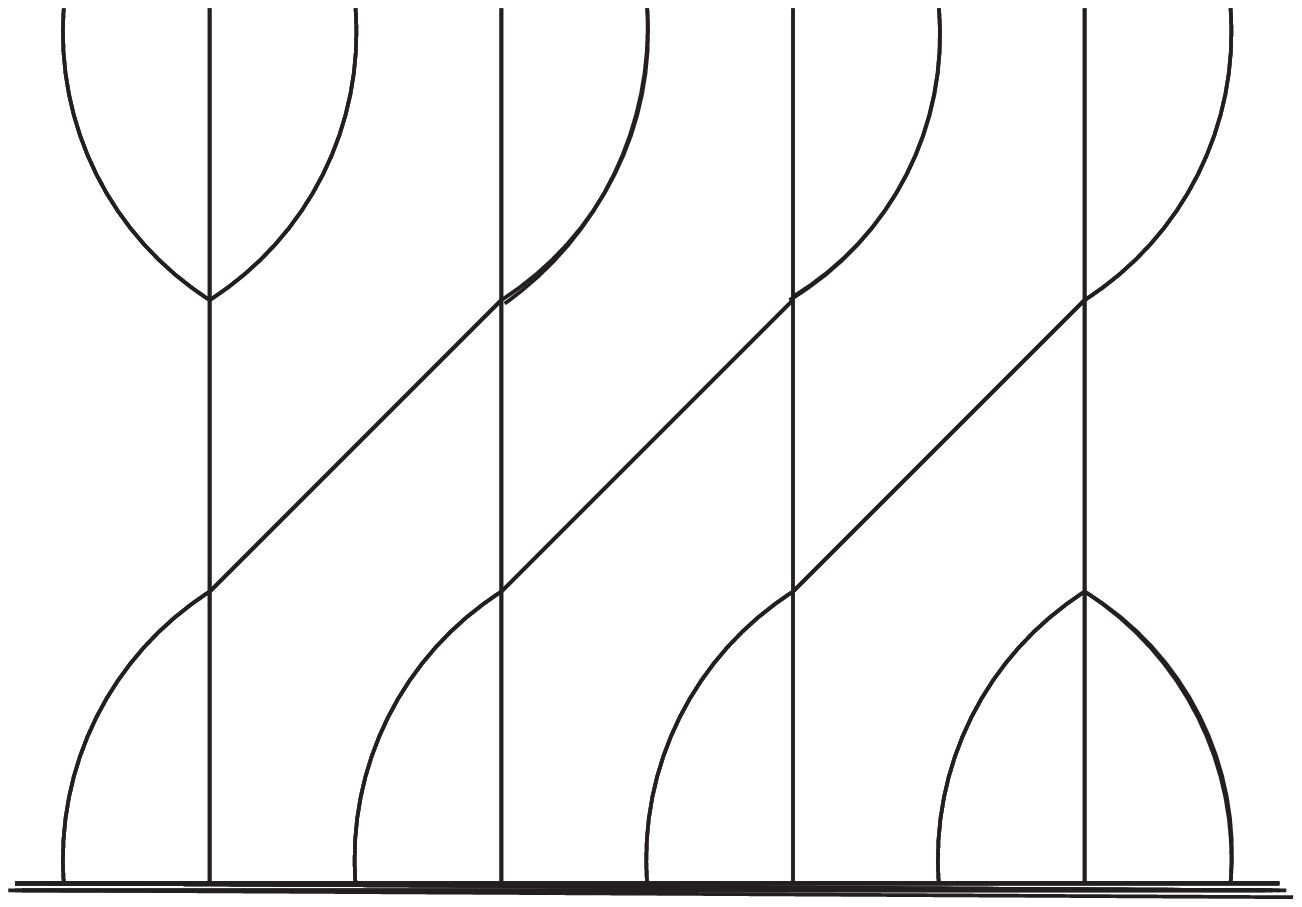}\\ $\mathcal{S}_e$}
\caption{Supergraphs corresponding to the maximal reshuffling
terms of the eight-loop dilatation operator of ABJM theory. From
left to right, and together with their reflected diagrams, they
are associated to each line in (\ref{dilatationmr}). The
horizontal bar represents the operator itself, or analogously, the
spin chain.\label{supergraphs}}
\end{figure}

We first rewrite the dilatation operator in the basis of
\emph{chiral functions} introduced in Appendix
\ref{PermutationStructures}, which is directly related to the chiral
structure of the supergraphs. The part of the dilatation operator
leading to maximal reshuffling of spins\footnote{We give here only
the terms which act non-trivially on even sites.} preserves the form of
(\ref{dilatationeven}):
\begin{equation}\label{dilatationmr}
 \begin{split}
  \mathcal{D}_{8,mr}=
             &\,m\,\chi(2,0,4,2)\\
             \vspace{0.4cm}
             &\,+l_3\,[\chi(0,4,2,6)+\chi(2,0,6,4)]\\
             &\,+\Big(-1+\frac{1}{2}m+i\,\epsilon_{2a}+\frac{1}{4}\beta\Big)\,[\chi(0,2,6,4)+\chi(0,6,4,2)]\\
             &\,+\Big(-1+\frac{1}{2}m-i\,\epsilon_{2a}+\frac{1}{4}\beta\Big)\,[\chi(2,0,4,6)+\chi(4,2,0,6)]\\
             \vspace{0.4cm}
             &\,-5\,(\chi(0,2,4,6)+\chi(6,4,2,0)]\, .
 \end{split}
\end{equation}
This part of the dilatation operator is sufficient to compute
the parameter $\beta_{2,3}^{(6)}$. $\mathcal{D}_{8,mr}$ has just
the same form as the maximal reshuffling hamiltonian of
$\mathcal{N}=4$ SYM \cite{Beisert:2007hz}. The coefficients of the
five terms in (\ref{dilatationmr}) can be computed from the
supergraphs contributing to $\chi(2,0,4,2)$, $\chi(0,4,2,6)$,
$\chi(0,2,6,4)$, $\chi(2,0,4,6)$, $\chi(0,2,4,6)$. Only one
supergraph contributes to each term and it contains only scalar
interactions. The coefficient of the last term was already fixed
by imposing the dispersion relation on one-magnon states. The
first four terms contain precisely four undetermined parameters,
namely $m$, $l_3$, $\epsilon_{2a}$ and $\beta$: in order to
compute such parameters we need to evaluate the supergraphs in
Figure \ref{supergraphs} and isolate the overall UV divergence by
subtracting all their UV subdivergences.

\begin{figure}
\centering
\parbox{2.5cm}{\centering\includegraphics[width=1.9cm, height=1.3cm]{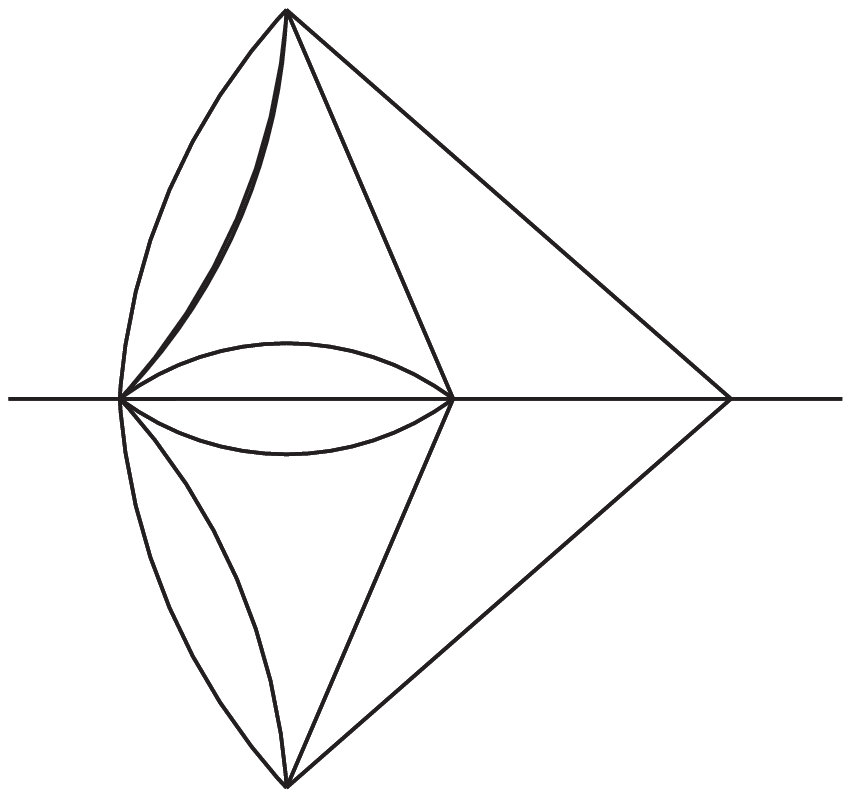}\\ $I_{8a}$}
\parbox{3cm}{\centering\includegraphics[width=2.5cm, height=1.2cm]{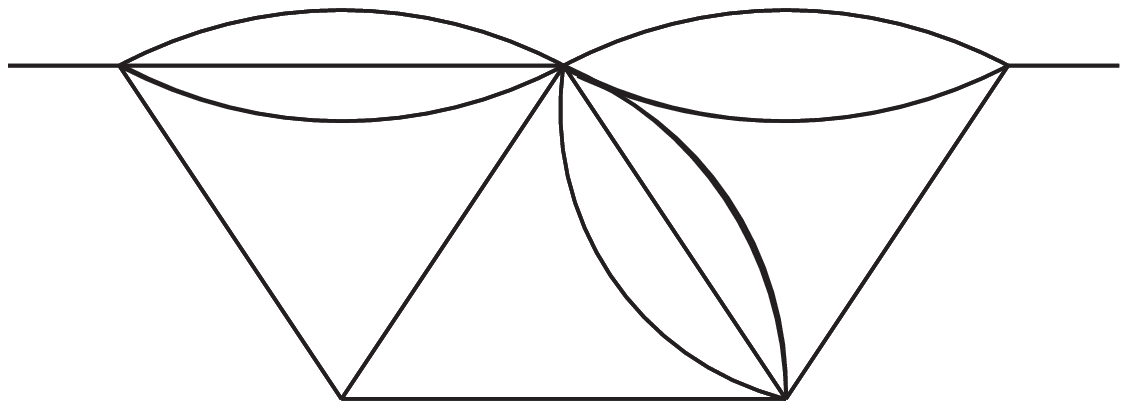}\\ $I_{8b}$}
\parbox{3cm}{\centering\includegraphics[width=2.5cm, height=1.2cm]{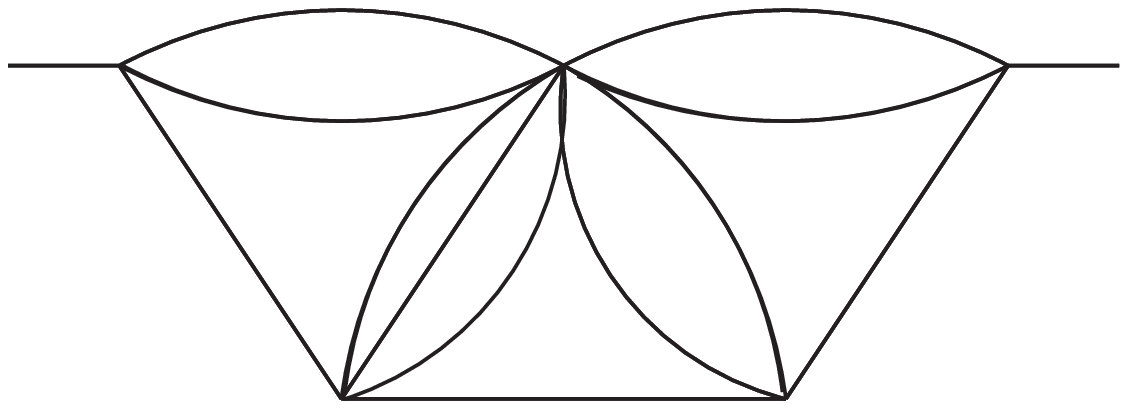}\\ $I_{8c}$}
\parbox{3cm}{\centering\includegraphics[width=2.5cm, height=1.2cm]{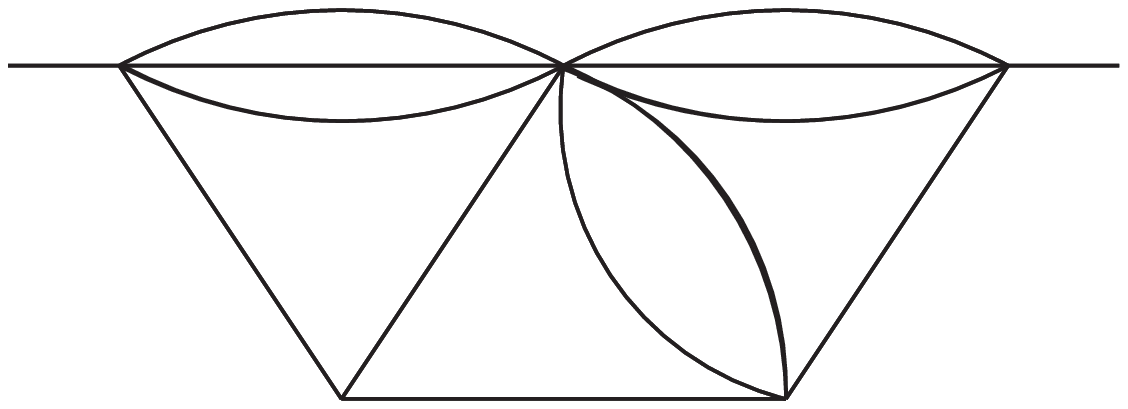}\\ $I_{8d}$}
\parbox{3cm}{\centering\includegraphics[width=2.5cm, height=1.2cm]{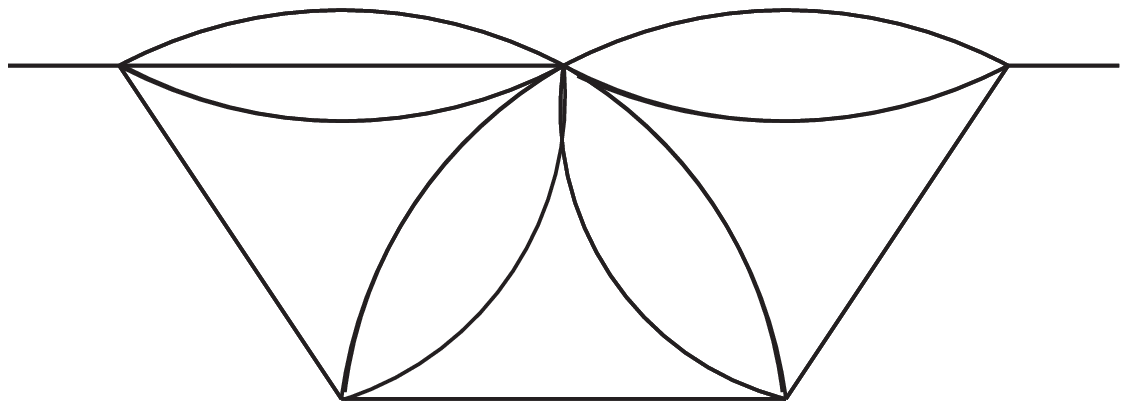}\\ $I_{8e}$}
\caption{Momentum-space integrals obtained from supergraphs
$\mathcal{S}_a$, $\mathcal{S}_b$, $\mathcal{S}_c$,
$\mathcal{S}_d$, $\mathcal{S}_e$ after completion of the D-algebra
procedure.\label{momentumintegrals}}
\end{figure}

First of all, we perform standard D-algebra manipulations (see
Appendix \ref{superspace}) to reduce the supergraphs to ordinary
momentum-space integrals $I_{8x}$. Such integrals are displayed in
Figure \ref{momentumintegrals}. The Feynman rules imply that the
color factor is always $N^8$, which combines with
$\left(\frac{4\pi}{k}\right)^8$ to give the right power of the 't
Hooft coupling, \emph{i.e.} $\lambda^8$ (the $(4\pi)^8$ factor
will be simplified by the factor $\frac{1}{(4\pi)^8}$ present in
each integral $I_{8x}$). The pole parts of the integrals with the
subdivergences subtracted are denoted by $\bar{I}_{8x}$. The
maximal reshuffling supergraphs, thus, evaluate to:

\begin{equation}
\mathcal{S}_a\;=\;\graph{Sa}{11}\,=\,\lambda^8\, (4\pi)^8\,
\bar{I}_{8a}|_{\frac{1}{\varepsilon}}\;\chi(2,0,4,2)\, ,
\end{equation}

\begin{equation}
\mathcal{S}_b\;=\;\graph{Sb}{10}\,=\,\lambda^8\, (4\pi)^8\,
\bar{I}_{8b}|_{\frac{1}{\varepsilon}}\;\chi(0,4,2,6)\, ,
\end{equation}

\begin{equation}
\mathcal{S}_c\;=\;\graph{Sc}{8}\,=\,\lambda^8\, (4\pi)^8\,
\bar{I}_{8c}|_{\frac{1}{\varepsilon}}\;\chi(0,2,6,4)\, ,
\end{equation}

\begin{equation}
\mathcal{S}_d\;=\;\graph{Sd}{8}\,=\,\lambda^8\, (4\pi)^8\,
\bar{I}_{8d}|_{\frac{1}{\varepsilon}}\;\chi(2,0,4,6)\, ,
\end{equation}

\begin{equation}
\mathcal{S}_e\;=\;\graph{Se}{8}\,=\,\lambda^8\, (4\pi)^8\,
\bar{I}_{8e}|_{\frac{1}{\varepsilon}}\;\chi(0,2,4,6)\, .
\end{equation}
The subtracted integrals are computed in Appendix
\ref{integrals} and are given by:
\begin{equation}\label{poles}
\bar{I}_{8a}\;=\;\;\;\graph{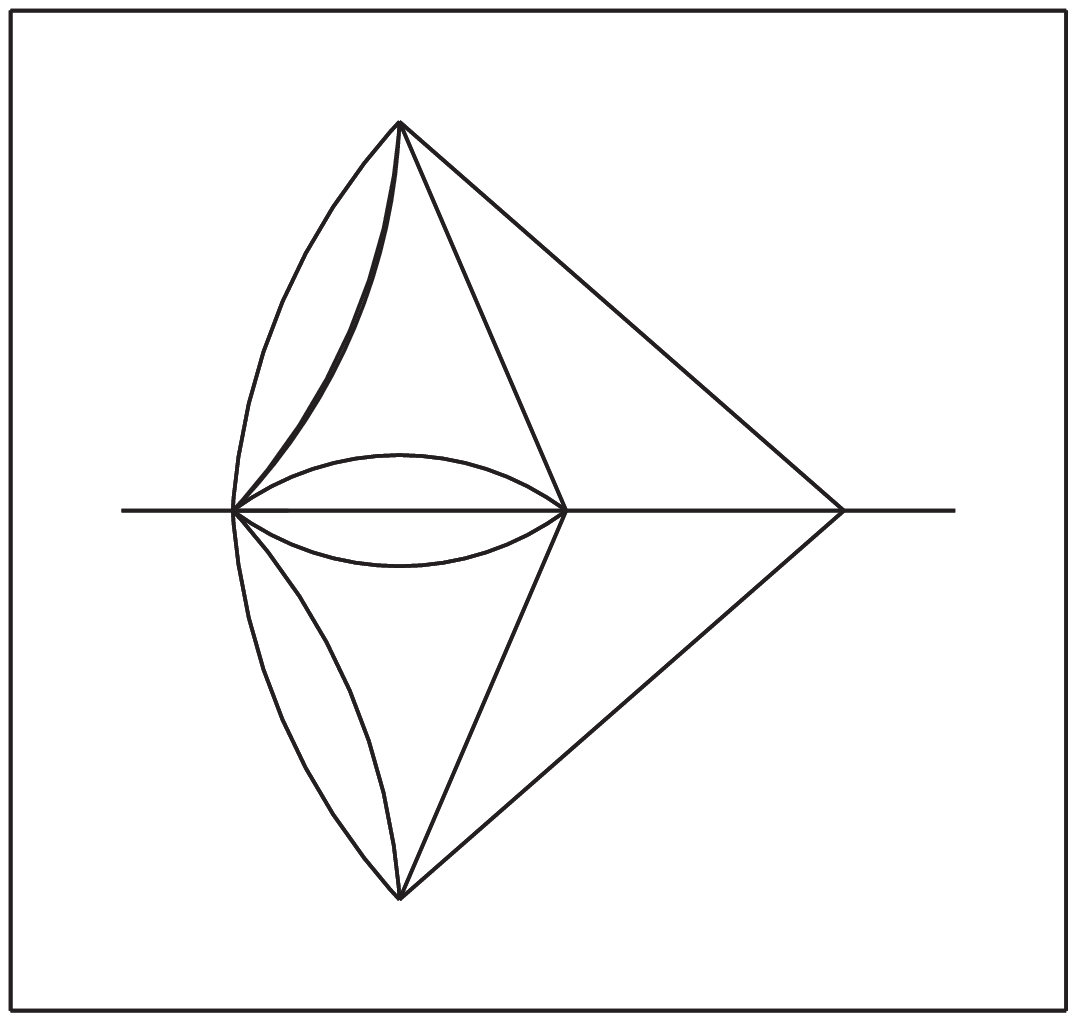}{10}\,\;\;\;=\;\;\;\frac{1}{(4\pi)^8}\,\left(-\frac{1}{3072\,\varepsilon^4}+\frac{1}{192\,\varepsilon^3}-\frac{1}{48\,\varepsilon^2}
+\frac{y}{\varepsilon}\right)\,
,
\end{equation}

\begin{displaymath}
\bar{I}_{8b}\;=\;\graph{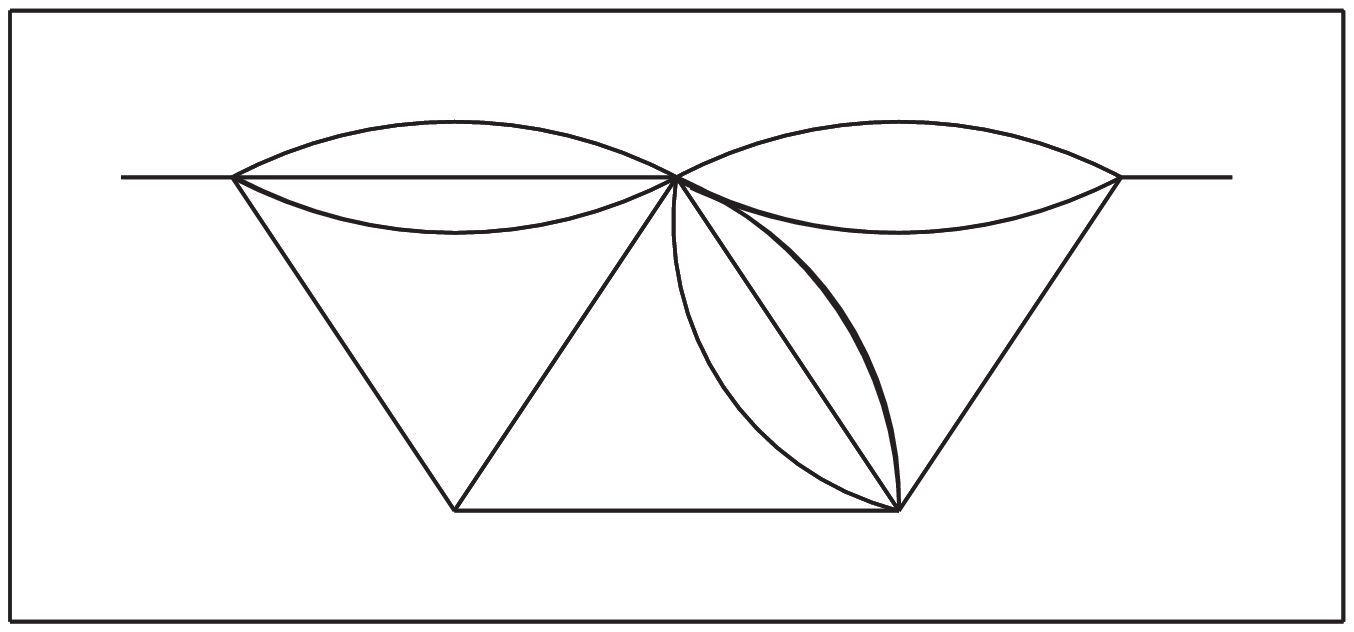}{6}\;=\;\frac{1}{(4\pi)^8}\left(-\frac{5}{6144\,\varepsilon^4}+\frac{5}{768\,\varepsilon^3}-\frac{1}{384\,\varepsilon^2}-\frac{1}{32\,\varepsilon}\right)\,
,
\end{displaymath}

\begin{displaymath}
\bar{I}_{8c}\;=\;\graph{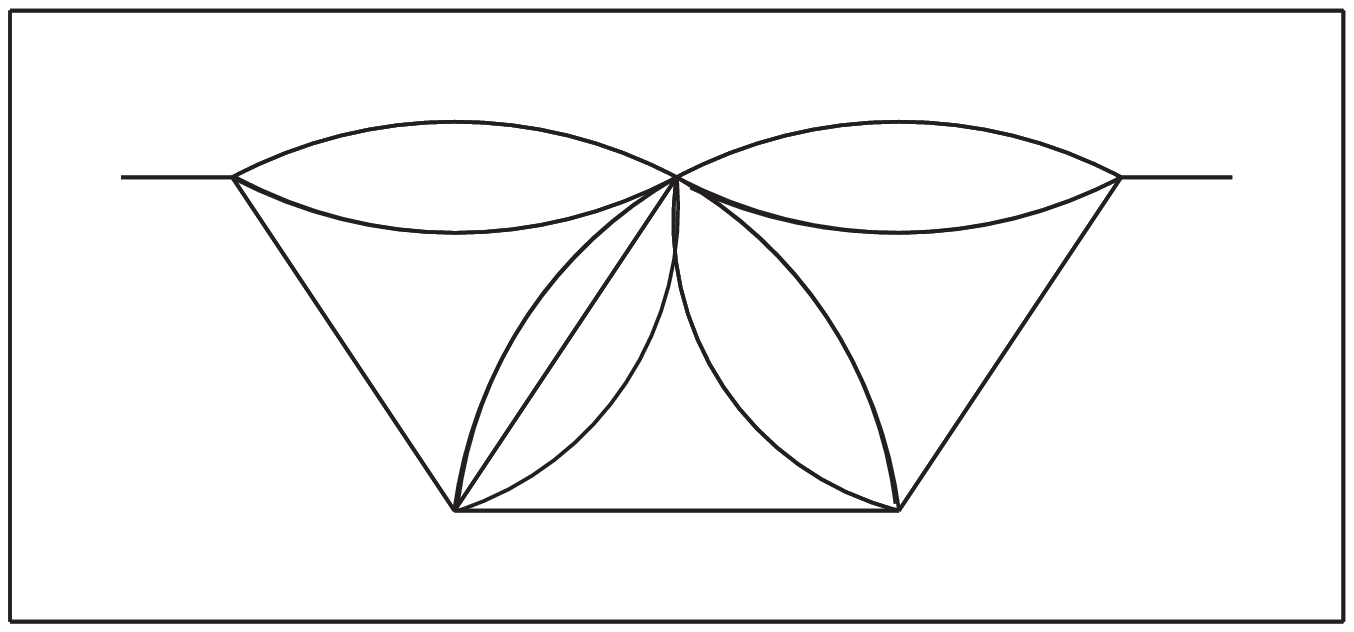}{6}\;=\;\frac{1}{(4\pi)^8}\left(-\frac{1}{2048\,\varepsilon^4}+\frac{1}{128\,\varepsilon^3}-\frac{3}{64\,\varepsilon^2}+\frac{5}{192\,\varepsilon}\right)\,
,
\end{displaymath}

\begin{displaymath}
\bar{I}_{8d}\;=\;\graph{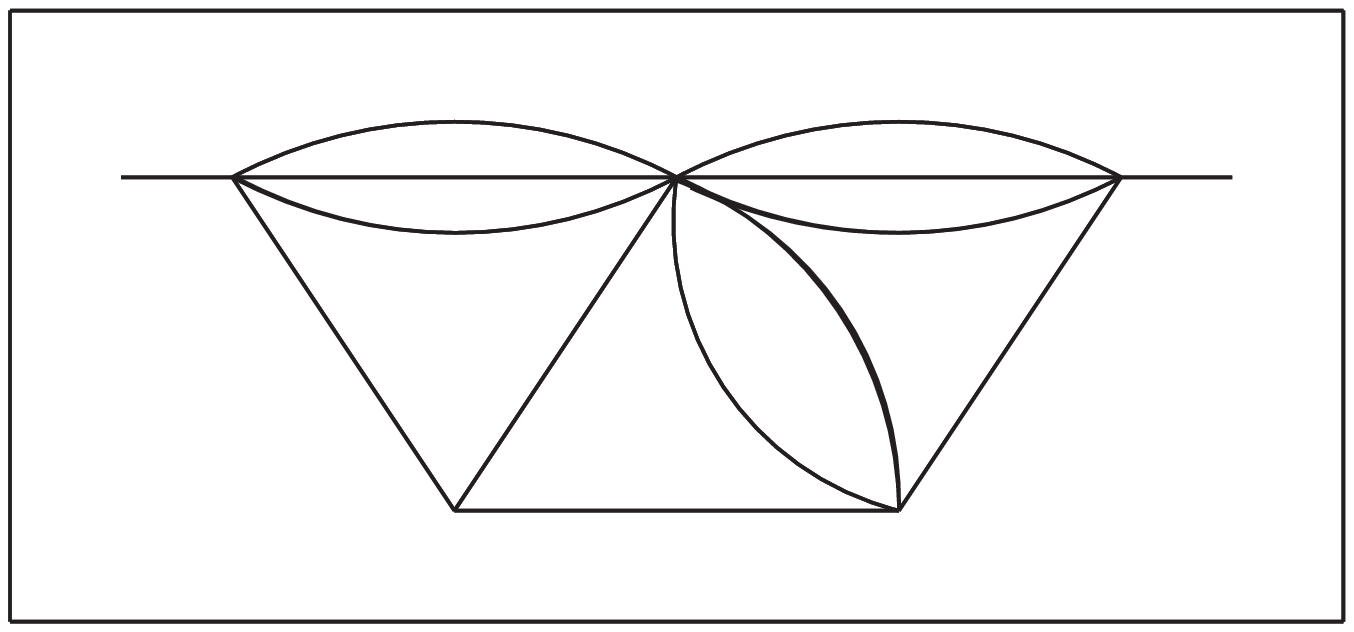}{6}\;=\;\frac{1}{(4\pi)^8}\left(-\frac{1}{2048\,\varepsilon^4}+\frac{1}{192\,\varepsilon^3}-\frac{1}{64\,\varepsilon^2}-\frac{11}{192\,\varepsilon}\right)\,
,
\end{displaymath}

\begin{displaymath}
\bar{I}_{8e}\;=\;\graph{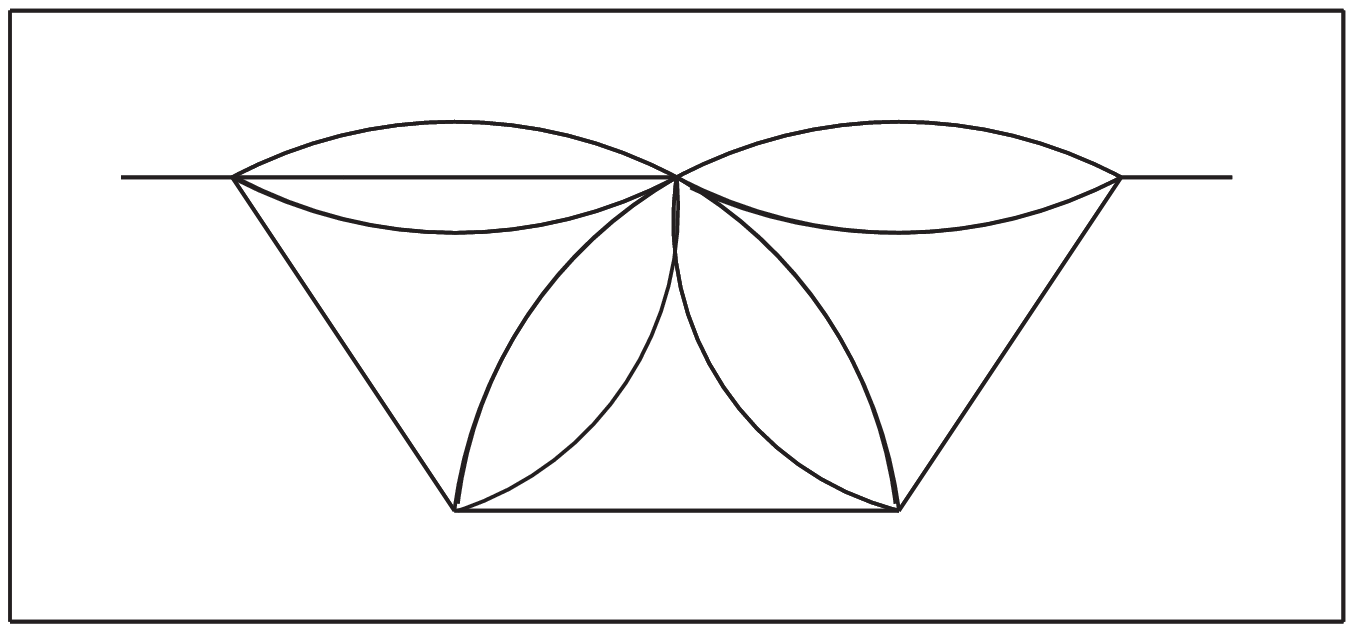}{6}\;=\;\frac{1}{(4\pi)^8}\left(-\frac{1}{6144\,\varepsilon^4}+\frac{1}{256\,\varepsilon^3}-\frac{19}{384\,\varepsilon^2}+\frac{5}{16\,\varepsilon}\right)\,
.
\end{displaymath}
\\For their evaluation, we used the \emph{Gegenbauer polynomials $x$-space
technique} (GPXT). We briefly review this technique in Appendix
\ref{GPXT}.

We could analytically compute all the integrals in (\ref{poles}) apart
from the $\frac{1}{\varepsilon}$ pole of $\bar{I}_{8a}$, denoted by $y$ in the
(\ref{poles}): in this case GPXT seems to be quite inefficient, so we followed
a different strategy. We evaluated the integral numerically through
Mellin-Barnes techniques. Remarkably, guided by the transcendentality principle, it was possible to extract
the corresponding analytical result from the numerical one by
means of the PSLQ algorithm \cite{PSLQ}. It is given by
\begin{equation}
 y=-\frac{5}{32}+\frac{1}{8}\,\zeta(3)\, .
\end{equation}
All the details are given in Appendix \ref{integrals}.

We can write now the final results for the supergraphs
contributing to maximal reshuffling terms of the dilatation
operator:
\begin{equation}\begin{split}
\mathcal{S}_a\;=&\;\lambda^8\,\left(-\frac{5}{32}+\frac{1}{8}\,\zeta(3)\right)\;\chi(2,0,4,2)\,
,\\
\mathcal{S}_b\;=&\;-\lambda^8\,\frac{1}{32}\;\chi(0,4,2,6)\, ,\\
\mathcal{S}_c\;=&\;\lambda^8\,\frac{5}{192}\;\chi(0,2,6,4)\, ,\\
\mathcal{S}_d\;=&\;-\lambda^8\,\frac{11}{192}\;\chi(2,0,4,6)\, ,\\
\mathcal{S}_e\;=&\;\lambda^8\,\frac{5}{16}\;\chi(0,2,4,6)\, ,\\
\end{split}\end{equation}
According to the definition (\ref{DZ}) of the dilatation operator, the 
corresponding contributions to the renormalization factor are
obtained multiplying these values by -16. At this point, we can
find the unknown parameters appearing in (\ref{dilatationmr}). Our
results are
\footnote{The presence of the factor $i$ in the unphysical
$\epsilon_{2a}$
seems to spoil hermiticity of the dilatation operator, see
\cite{Beisert:2007hz} for further comments on this. Moreover, the
coefficient of the chiral function $\chi(0,2,4,6)$ turns out to be
-5, as it was already known from the previous section: it, thus,
constitutes a non trivial check of the whole procedure.}:
\begin{equation}\label{results}
m\,=\,\frac{5}{2}-2\zeta(3)\,
,\;\;\;\;\;\;\;\;l_3\,=\,\frac{1}{2}\,
,\;\;\;\;\;\;\;\;\beta\,=\,4\,\zeta(3)\, ,\;\;\;\;\;\;\;\;\epsilon_{2a}\,=\,\frac{2}{3}i\, .\\
\end{equation}
In particular we have computed the value of the leading order
coefficient of the dressing phase, which is
$\beta_{2,3}^{(6)}=4\zeta(3)$. It is the same value found in
$\mathcal{N}=4$ SYM for the $\beta_{2,3}^{(3)}$ coefficient. In
the Conclusions we comment on this result.

\section{Conclusions}
In this paper we have investigated the dressing phase of ABJM theory
at weak coupling. It appears in the Bethe equations, as well
as in the dilatation operator, starting at eight loops. A simple
procedure based on the Bethe Ansatz allowed us to construct the
asymptotic dilatation operator in the $SU(2)\times SU(2)$ sector
up to six loops. We
verified that the dressing phase is not present up to this
order. Furthermore, we could largely constrain the eight-loop
dilatation operator and show that the two $SU(2)$ sectors are coupled starting at this order. 
In this case we focused on the terms corresponding to maximal interactions: they contain the
unknown parameter $\beta_{2,3}^{(6)}$, which is the leading order
coefficient of the dressing phase. Thanks to superspace techniques
and with the help of the transcendentality principle, we could
directly compute its value from Feynman supergraphs. A great
simplification follows from the fact that we considered maximal
interactions: we needed to compute a very small number of Feynman
diagrams (in fact, just three are necessary to completely fix
$\beta_{2,3}^{(6)}$) which contain only scalar interactions. The
result we found is $\beta_{2,3}^{(6)}=4\zeta(3)$ and coincides
with the value of the leading coefficient of the dressing phase of
$\mathcal{N}=4$ SYM.  We note that from a diagrammatic point of view the reason for this coincidence is not manifest since we are dealing with integrals in different dimensions.  In the  $\mathcal{N}=4$ SYM case it was argued
\cite{Beisert:2006ez} that the weak-coupling expansion of
$\beta_{r,s}(\lambda)$ can be extrapolated by analytical
continuation of the all-loop strong-coupling coefficient functions
 \cite{Hernandez:2006tk,Beisert:2006ib} $\beta_{r,s}(g)=\sum_{n=0}^\infty
 c_{r,s}^{(n)}g^{r+s-n-1}$, with
\begin{equation}
c_{r,s}^{(n)}=\frac{(1-(-1)^{r+s}\zeta(n))}{2(-2\pi)^n\Gamma(n-1)}(r-1)(s-1)\frac{\Gamma[(\frac{1}{2}(s+r+n-3)]\Gamma[(\frac{1}{2}(s-r+n-1)]}{\Gamma[(\frac{1}{2}(s+r-n+1)]\Gamma[(\frac{1}{2}(s-r-n+3)]}\,
,
\end{equation}
to negative values of $n$:
\begin{equation}
\beta_{r,s}(g)=-\sum_{n=1}^{\infty}c_{r,s}^{(-n)}\,g^{r+s+n-1}\, ,
\end{equation}
where $g=\frac{\sqrt{\lambda}}{4\pi}$. So, the dressing phase
coefficients were conjectured to be
\begin{equation}
\beta_{r,s}^{(\ell)}=-c_{r,s}^{(r+s-2\ell-1)}\, .
\end{equation}
In particular, $c_{2,3}^{(-2)}=-4\zeta(3)$, so that the leading order coefficient of the SYM dressing
phase was $\beta_{2,3}^{(3)}=4\zeta(3)$. It was shown that this
guess is indeed correct \cite{Beisert:2007hz}.

Since  in the Bethe Ansatz proposal of \cite{Gromov:2008qe} the strong coupling limit of the dressing phase is the same of the $AdS_5/\text{CFT}_4$ case, we can suppose, by analogy of the $\mathcal{N}=4$ SYM case and
with the usual replacement $g\rightarrow h(\lambda)$, that in ABJM
we have
\begin{equation}\label{conjecture}
\beta_{r,s}(\lambda)=-\sum_{n=1}^{\infty}c_{r,s}^{(-n)}\,h(\lambda)^{r+s+n-1}\,
.
\end{equation}
For $r=2$, $s=3$, we see that we should obtain (since $c_{2,3}^{(-1)}=c_{2,3}^{(-3)}=0$)

\begin{equation}
 \beta_{2,3}(\lambda)=- h_2^3 c_{2,3}^{(-2)}\,\lambda^6-\left(h_2^{4} c_{2,3}^{(-4)}+3 h_4h_2^2 c_{2,3}^{(-2)}\right)\,\lambda^8+\cdots
\end{equation}
We observe that, since the first weak coupling coefficient of
$h(\lambda)$ is $h_2=1$, the leading order contribution to ABJM
dressing phase is just $\beta_{2,3}^{(6)}=4\zeta(3)$, which is
precisely the same value obtained in $\mathcal{N}=4$ SYM. At
higher orders, however, the $c_{r,s}^{(-n)}$ coefficients mix with
the non trivial coefficients of $h(\lambda)$ yielding dressing
phase coefficients which differ from the $\mathcal{N}=4$ SYM ones.
We stress that, in our context, this is merely a conjecture:
anyway, we computed the coefficient $\beta_{2,3}^{(6)}$ by
independent field theory techniques and we found that the above
prediction is indeed correct. It would be interesting to further
test the validity of (\ref{conjecture}) beyond the leading order,
through direct computations: hopefully, restricting to the case of
maximal interactions, the calculations will be simplified and have
a chance of being performed with standard field theory techniques.

\acknowledgments
S.S. would like to thank F. Fiamberti for useful discussions.
This work has been supported in part by INFN and MIUR-PRIN contract
2009-KHZKRX.  

\newpage

\appendix

\section{ABJM Theory in $\mathcal{N}=2$
superspace}\label{superspace} In this Appendix we briefly review
the $\mathcal{N}=2$ superspace formulation of ABJM theory. This
was first given in \cite{Benna:2008zy}, but we follow the
notations used in \cite{Leoni:2010tb} which are adapted from the
ones of \cite{Gates:1983nr}. ABJM theory has two $\mathcal{N}=2$
vector supermultiplets, $V$ and $\hat{V}$, with $V$ transforming
in the adjoint of the first $U(N)$ factor and $\hat{V}$ in the
adjoint of the second $U(N)$ factor of the gauge group. In order
to extend the supersymmetry to $\mathcal{N}=6$, the ABJM action
also contains two sets of chiral matter superfields, $Z^A$ and
$W_A$ with $A=1,2$. $Z^A$ and $W_A$ transform respectively in the
$({\bf 2}, {\bf 1})$ and $({\bf 1}, {\bf 2})$ of the global
$SU(2)\times SU(2)$ flavour group. Moreover, they transform in the
bifundamental representations $({\bf N}, {\bf \bar{N}})$ and
$({\bf \bar{N}}, {\bf N})$ of $U(N)\times U(N)$. The gauge fixed
ABJM action in $\mathcal{N}=2$ superspace reads
 \begin{equation}\label{action}\begin{split}
S=&
\frac{k}{4\pi}\;\left\{\int\text{d}^3x\,\text{d}^4\theta\,\int_0^1
\text{d}t\; \text{Tr}\left(V\,\bar{\text{D}}^\alpha
e^{-tV}\text{D}_\alpha e^{tV}-\hat{V}\,\bar{\text{D}}^\alpha
e^{-t\hat{V}}\text{D}_\alpha e^{t\hat{V}}\right)\right.\\
&+\int\text{d}^3x\,\text{d}^4\theta\,\text{Tr}\left(\bar{Z}_A
e^{V} Z^A e^{-\hat{V}}+\bar{W}^B e^{\hat{V}}W_B
e^{-V}\right)\\
&+\frac{i}{2}\left.\left[\int\text{d}^3x\,\text{d}^2\theta\,
\epsilon_{AC}\epsilon^{BD}\,\text{Tr}Z^AW_BZ^CW_D+\int\text{d}^3x\,\text{d}^2\bar{\theta}\,
\epsilon^{AC}\epsilon_{BD}\,\text{Tr}\bar{Z}_A\bar{W}^B\bar{Z}_C\bar{W}^D\right]\right.
\\
&+\;\text{gauge fixing and ghost terms}\bigg\}\, .
\end{split}\end{equation}
The first line contains the Chern-Simons action, the second line
contains the kinetic term of the matter fields and their coupling
with gauge fields, while the third line is the superpotential.

The three-dimensional, $\mathcal{N}=2$ superspace spinor covariant
derivatives $\text{D}_\alpha,\bar{\text{D}}_\alpha $ satisfy the
algebra
\begin{equation}
\begin{split}
\{\text{D}_\alpha,\text{D}_\beta\}&=\{\bar{\text{D}}_\alpha,\bar{\text{D}}_\beta\}=0\,
,\;\;\;\;\{\text{D}_\alpha,\bar{\text{D}}_\beta\}=p_{\alpha\beta}\,
.
\end{split}
\end{equation}
The metric $\epsilon_{AB}$ for the SU(2) flavor indices is given
by
\begin{equation}
\begin{aligned}
\epsilon_{12}=1\, ,\;\;\;\;\;\; \epsilon^{12} =1\, ,\;\;\;\;\;\;
\epsilon^{AB}\epsilon_{CD}=\delta^A_C\delta^B_D-\delta^A_D\delta^B_C\,
.
\end{aligned}
\end{equation}
For the integration over the superspace our conventions are
$\int\d^2\theta=\frac{1}{2}\partial^\alpha\partial_\alpha$,
$\int\d^2{\bar\theta}=\frac{1}{2}{\bar\partial}^\alpha{\bar\partial}_\alpha$
and $\int\d^4{\theta}=\int\d^2{\theta}\d^2{\bar\theta}$, such that
\begin{equation}
\begin{aligned}
&\int\d^3x\d^2\theta=\int\d^3x\,\text{D}^2|_{\theta={\bar\theta}=0}
~,~~~
\int\d^3x\d^2{\bar\theta}=\int\d^3x\,{\bar{\text{D}}}^2|_{\theta={\bar\theta}=0}~,~~~
\\
&\int\d^3x\d^4\theta=\int\d^3x\,\text{D}^2{\bar{\text{D}}}^2|_{\theta={\bar\theta}=0}
~.
\end{aligned}
\end{equation}
The $\theta$-space $\delta$-function is given by
\begin{equation}
\delta^4(\theta-\theta')=(\theta-\theta')^2({\bar\theta}-{\bar\theta}')^2
~.
\end{equation}
We now give the Euclidean Feynman rules (\emph{i.e.} we have Wick-rotated to $e^{S}$ in the path integral) of
the theory, relevant for the computations of the diagrams in
Figure \ref{supergraphs}. The chiral field propagators are
\begin{equation}
\graph{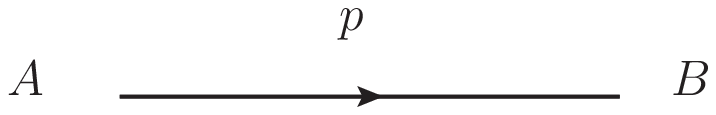}{3}\;=\;\langle  Z^B(p)\bar Z_A(-p)\rangle
=\,\langle \bar W^B(p)W_A(-p)\rangle
=\,\frac{4\pi}{k}\frac{\delta_A^B}{p^2}\delta^4(\theta_1-\theta_2)\,
,
\end{equation}
where diagonality in the gauge indices have been suppressed. The
vertices are obtained by taking the functional derivatives of the
Wick rotated action w.r.t. the corresponding superfields. When a
functional derivatives w.r.t. the (anti)-chiral superfields is
taken, factors of ($\text{D}^2$) $\bar{\text{D}}^2$ are generated
in the vertices. We need only the quartic superpotential vertices:
\begin{equation}
\graph{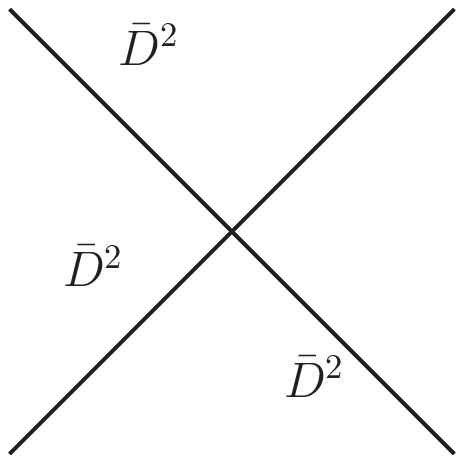}{8}\;=\;i\epsilon^{AC}\epsilon_{BD}\,\frac{k}{4\pi}\,\big[
\text{Tr}\big(B^{\underline{a}}B_{\underline{b}}B^{\underline{c}}B_{\underline{d}}\big)
-\text{Tr}\big(B^{\underline{c}}B_{\underline{b}}B^{\underline{a}}B_{\underline{d}}\big)
\big]
\end{equation}
\begin{equation}
\graph{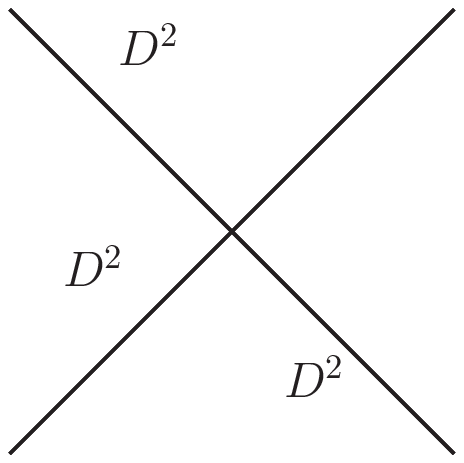}{8}\;=\;i\epsilon_{AC}\epsilon^{BD}\,\frac{k}{4\pi}\big[
\text{Tr}\big(B_{\underline{a}}B^{\underline{b}}B_{\underline{c}}B^{\underline{d}}\big)
-\text{Tr}\big(B_{\underline{c}}B^{\underline{b}}B_{\underline{a}}B^{\underline{d}}\big)
\big]\, ,
\end{equation}
where the color indices are labeled $(a, b, c, d)$
counter-clockwise starting with the leg in the upper left corner.
Note also that, in a standard way, one of the ($\text{D}^2$)
$\bar{\text{D}}^2$ factors has been absorbed into the (anti)chiral
integration such that the integration measure of the (anti)chiral
vertex is promoted to the full superspace measure. We have
introduced matrices $B^{\underline{a}}$ and $B_{\underline{a}}$,
with underlined $\underline{a}=1,\cdots,N^2$ indices that
transform in the $({\bf N}, {\bf \bar{N}})$ of the gauge group
$U(N)\times U(N)$.

\section{Permutation structures}\label{PermutationStructures}
The spin-chains which arise in the study of ABJM theory at $\ell$
loops in the $SU(2)\times SU(2)$ sector are long-range
deformations of the alternating Heisenberg spin chain with
next-to-nearest neighbor interactions. The interactions among
spins are represented by products of permutations of
next-to-neighbor spins, therefore we introduce the
\emph{permutation structures} as \cite{Minahan:2009aq}
\begin{equation}
 \{a_1,\ldots,a_n\}=\sum_{j=1}^L\,P_{2j+a_1,2j+a_1+2}\cdots
 P_{2j+a_n,2j+a_n+2}\, ,
\end{equation}
where the permutation operators $P_{i,j}$ exchange spins at
sites $i$ and $j$. Indices are understood modulo $2L$. If $n=0$,
we denote $\{\,\}=L$. The range $\mathcal{R}$ of a permutation structure is
\begin{equation}
\mathcal{R}=\max(a_1,\ldots,a_n)-\min(a_1,\ldots,a_n)+3\, .
\end{equation}
The integer $n$ is called \emph{length} of the permutation
structure and, in terms of the associated Feynman diagrams, it
coincides with the number of chiral and antichiral vertices. The
permutation structures satisfy the following relations:
\begin{equation}\begin{split}
                    \{\ldots,a,a,\ldots\}&=\{\ldots,\ldots\}\, ;\\
                    \{\ldots,a,b,\ldots\}&=\{\ldots,b,a,\ldots\}\, ,\;\;\;\;\;\;\;\text{if
}|b-a|\neq 2\, ;\\
                    \{a,b,\ldots\}&=\{a+2m,b+2m,\dots\}\, ,\;\;\;\;\;\;\; m=0,1,2,\ldots\\
                    \{\ldots,a,a+2,a,\ldots\}&=\{\ldots,a,a+2,\ldots\}+\{\ldots,a+2,a,\ldots\}-\{\ldots,a,\ldots\}\\
                                  &\phantom{=}\,-\{\ldots,a+2,\ldots\}+\{\ldots,\ldots\}\,
                                  .
                   \end{split}
\end{equation}
Under hermitian conjugation and parity, they transform as
\begin{equation}\begin{split}
                                  \{a_1,\ldots,a_n\}^\dagger
                                  &=\{a_n,\ldots,a_1\}\, ;\\
                                  \mathcal{P}\{a_1,\ldots,a_n\}\mathcal{P}^{-1}&=\{-a_1,\ldots,-a_n\}\,
                                  .
                   \end{split}
\end{equation}
The permutation structures represent a convenient basis
for the dilatation operator. Nevertheless, when dealing with
Feynman diagrams, it is useful to use a different basis, namely
the basis of \emph{chiral functions}, defined in terms of
permutation structures as \cite{Fiamberti:2007rj, 
Fiamberti:2008sh, Fiamberti:2010ra}
\begin{equation}\begin{split}
\chi()=&\,\{\,\}\, ,\\
\chi(a)=&\,\{a\}-\{\,\}\, ,\\
\chi(a,b)=&\,\{a,b\}-\{a\}-\{b\}+\{\,\}\, ,\\
\chi(a,b,c)=&\,\{a,b,c\}-\{a,b\}-\{a,c\}-\{b,c\}+\{a\}+\{b\}+\{c\}-\{\,\}\,
,\\
\chi(a,b,c,d)=&\,\{a,b,c,d\}-\{a,b,c\}-\{a,b,d\}-\{a,c,d\}-\{b,c,d\}\\
&\,+\{a,b\}+\{a,c\}+\{a,d\}+\{b,c\}+\{b,d\}+\{c,d\}\\
&\,-\{a\}-\{b\}-\{c\}-\{d\}+\{\,\}\, .
\end{split}\end{equation}
These functions are directly related to the chiral structure of
a supergraph and precisely describe the flavor flow of fields
inside (\ref{su2operators}) under the action of the interaction.
All diagrams contributing to the same chiral function share the
same chiral structure. For the maximal interactions encountered in
this work, no vector interactions can be present and the chiral
structure alone determines the unique diagram which contributes.
The chiral structure of the function $\chi(a_1,\ldots,a_n)$
contains $n$ chiral vertices and $n$ antichiral vertices,
connected by $\langle Z\bar{Z}\rangle$ or $\langle
W\bar{W}\rangle$ propagators. Each pair of chiral and antichiral
vertices (plus the propagator connecting them) describes an
elementary permutation of next-to-nearest neighbors.

\section{The Gegenbauer Polynomial $x$-Space
Technique}\label{GPXT} In this Appendix we review the Gegenbauer
Polynomial $x$-Space Technique (GPXT), which has been applied to
find the UV divergent part of the Feynman integrals needed in the
present work. The technique was introduced in
\cite{Chetyrkin:1980pr} and developed in
\cite{Kotikov:1995cw,Kotikov:2001sd}. See also
\cite{SiegGPXT,Fiamberti:2010ra} for useful reviews. All the
integrals are computed using dimensional regularization in
Euclidean space of dimension
\begin{equation}
D=2(\lambda+1)\, ,
\end{equation}
where
\begin{equation}\label{lambda}
\lambda=\frac{1}{2}-\varepsilon
\end{equation}
in order to get three-dimensional space in the
$\varepsilon\rightarrow 0$ limit.
\\In our computation, a generic Wick-rotated $\ell$-loop integral in momentum space has always the form
\begin{equation}\label{loopint}
I_\ell=\frac{1}{(2\pi)^{\ell D}}\,\int\frac{d^Dk_1\cdots
d^Dk_\ell}{\Pi_1\cdots \Pi_P}\, ,
\end{equation}
where $P$ is the number of propagators $\Pi_i$, which, in general, depend on
the loop momenta $k_1,\ldots,k_\ell$ and the external momenta
$p_1,\ldots,p_e$. We always consider massless propagators.

A propagator with weight $\alpha$ in momentum space is
Fourier-transformed to coordinate space according to the following
formula
\begin{equation}\label{Fourier}
\frac{1}{k^{2\alpha}}=\frac{\Gamma(\lambda+1-\alpha)}{\Gamma(\alpha)\pi^{\lambda+1}}\,\int\frac{\di^Dx\,
e^{2ikx}}{x^{2(\lambda+1-\alpha)}}\, ,
\end{equation}
where $\Gamma(z)$ is the Euler gamma function. The weight $\alpha$ is a generic complex number.
We stress the presence of the unconventional factor 2 in the
exponential. According to this definition, we have
\begin{equation}\label{delta}
 \int \di^Dk\, e^{2ikx}=\pi^{2(\lambda+1)}\delta(x)\, .
\end{equation}
According to GPXT technique, the computations are made directly in coordinate space rather than in
momentum space. The technique is grounded on the observation that,
in $x$-space, the scalar propagator always depends on the
difference of two points,
\begin{equation}
\Delta(x_i,x_j)=\frac{1}{(x_i-x_j)^{2\lambda}}\, ,
\end{equation}
and can thus be expanded in terms of the Gegenbauer polynomials, which form an orthogonal set on the unit sphere in $\mathds{R}^D$. For the moment, $D$ is an arbitrary integer dimension and $\lambda=D/2-1$. The analytic continuation to complex $D$ will be done in a second step.

The Gegenbauer polynomials $C_n^\lambda$ are defined in terms of a
generating function,
\begin{equation}\label{generating}
 \frac{1}{(1-2xt+t^2)^\lambda}=\sum_{n=0}^\infty=C_n^\lambda(x)\,t^n\, ,
\end{equation}
where $x\in [-1,1]$. We refer to the quantity $\lambda$ as the weight of the polynomial, while $n$ is its index.
The Gegenbauer polynomials are orthogonal with respect to the
weight function $(1-x^2)^{\lambda-1/2}$:
\begin{equation}\label{orth}
 \int_{-1}^1\di x \,(1-x^2)^{\lambda-\frac{1}{2}}C_n^{\lambda}(x)C_m^{\lambda}(x)=\frac{\pi 2^{1-2\lambda}\Gamma(n+2\lambda)}{n! (n+\lambda) \Gamma(\lambda)^2}\,\delta_{nm}\, .
\end{equation}
The following particular values are often needed:
\begin{equation}\begin{split}
 C_n^\alpha(1)=&\frac{\Gamma(n+2\alpha)}{n!\Gamma(2\alpha)}\, ,\\
 &\phantom{}\\
C_0^\alpha(x)=&\, 1\, .
\end{split}\end{equation}

Let's now turn to the description of the technique. We will
consider here only diagrams with a single external momentum $p$,
entering and leaving the graph at points $x_{\text{in}}$ and
$x_{\text{out}}$ in $x$-space. The generic $\ell$-loop integral in
(\ref{loopint}) becomes, in $x$-space,
\begin{equation}\label{loopintx}
 I_{\ell}=\frac{\Gamma(\lambda)^P}{(4^\ell \pi^P)^{\lambda+1}}\,\int\di^Dx_1\cdots \di^Dx_{P-\ell}\frac{e^{2ip(x_{\text{out}}-x_{\text{in}})}}{\Delta_1\cdots\Delta_P}
\end{equation}
where $\Delta_i$ are the propagators in coordinate space.

It is now convenient to move to spherical coordinates in $D$
dimensions: to this purpose we define
\begin{equation}
 r=x^2\, ,\;\;\;\;\;\;\hat{x}=\frac{x}{\sqrt{r}}\, .
\end{equation}
So $r$ is the (squared) radial coordinate and $\hat{x}$ is the unit vector pointing in the same direction as $x$. The integration measure changes to
\begin{equation}
 \di^D x=\frac{1}{2}S_{D-1}r^\lambda\di r\di \hat{x}\, ,
\end{equation}
where
\begin{equation}
 S_{D-1}=\frac{2\pi^{\lambda+1}}{\Gamma(\lambda+1)}
\end{equation}
is the surface of the unit sphere in $\mathds{R}^D$. The integral (\ref{loopintx}), therefore, transforms to
\begin{equation}
 I_{\ell}=N_\lambda(\ell,P)\,\int\frac{\di r_1\cdots \di r_{P-\ell}\di \hat{x}_1\cdots \di \hat{x}_{P-\ell}r_1^{\lambda}\cdots r_{P-\ell}^\lambda\,e^{2ip(x_{\text{out}}-x_{\text{in}})}}{\Delta_1\cdots\Delta_P}
\end{equation}
with the normalization factor
\begin{equation}\label{normalization}
 N_\lambda(\ell,P)=\frac{\Gamma(\lambda+1)^\ell}{(4\pi)^{\ell(\lambda+1)}\lambda^P}\, .
\end{equation}
At this point we can expand the propagators in terms of the Gegenbauer polynomials: from (\ref{generating}) we have
\begin{equation}\label{propGeg}
 \Delta(x_i,x_j)=\frac{1}{(x_i-x_j)^{2\lambda}}=\frac{1}{M_{i,j}^\lambda}\,\sum_{n=0}^\infty C_n^\lambda(\hat{x}_i\cdot \hat{x}_j)\left(\frac{m_{i,j}}{M_{i,j}}\right)^{\frac{n}{2}}\, ,
\end{equation}
where we have introduced the notation
\begin{equation}
   m_{i,j}=\min(r_i,r_j)\, ,\;\;\;\;\;\;M_{i,j}=\max(r_i,r_j)\, .
\end{equation}
If we are interested only in the UV divergent part of
the loop integral, as in all the cases in this work, a great
simplification occurs: in $x$-space, UV divergences appear when all coordinates are small, so we can approximate the
exponential factor with unit, neglecting it. Since it is equivalent
to set to zero the external momentum $p$, dropping the exponential
will introduce IR divergences, in the region
where some coordinates are large, which mix to UV ones altering the
final result. We thus regulate them by
introducing an infrared cutoff $R$ in the radial
integrations. At the end of the computation, when all the
subdivergences have been subtracted, the principal part of the
integral must be independent of the regulator $R$, while the
finite part in general depends on $R$ and should be discarded from
the result.

The expansion of the propagators in series of Gegenbauer
polynomials allows us to separate the integral in radial and
angular parts and introduces as many infinite sums as the number
of the propagators themselves. However, we can minimize the number
of series in this way: thanks to translational invariance of the
integral (\ref{loopintx}), we can choose one of the vertices of
the diagram as the origin of $D$-dimensional space. We call this
vertex the \textit{root vertex}. All the propagators directly
connected to the root vertex, which we call \textit{root
propagators}, are simply given by $1/r_i^\lambda$, where $r_i$ is
the radial coordinate of the vertex which connects the propagator
to the root vertex, and so they don't produce any series expansion
in the Gegenbauer polynomials. Obviously, the best choice of the
root vertex is usually such that the number of propagators
attached to it is maximized. Therefore, in most cases, the root
vertex will coincide with the composite operator.

Once we have chosen the root vertex and we have expanded the
propagators depending on differences of coordinates in terms of
Gegenbauer polynomials, the angular and radial integrations are
performed separately.

The angular integration can be performed by repeated use of the
orthogonality relation (\ref{orth}) of the Gegenbauer polynomials,
which can be rewritten in terms of the angular variables as
\begin{equation}\label{orth1}
 \int\di\hat{x}\,C_n^\lambda(\hat{x}_i\cdot\hat{x})C_m^\lambda(\hat{x}\cdot\hat{x}_j)=\,\frac{\lambda}{n+\lambda}\,\delta_{nm}\,C_n^\lambda(\hat{x}_i\cdot\hat{x}_j)\, .
\end{equation}
The angular integration is normalized as $\int\di\hat{x}=1$. In particular, since $C_0^\lambda(x)=1$, we have
\begin{equation}
 \int\di\hat{x}_i\di\hat{x}_j\,C_n^\lambda(\hat{x}_i\cdot\hat{x}_j)=\,\delta_{n0}\, .
\end{equation}

Kr\"{o}necker deltas can be used to decrease the number of
summations. Angular integration is usually the hardest part in
multiloop computations and can be extremely simplified if the root vertex 
is chosen so as to minimize the angular loop number and the number
of infinite summations. In fact, GPXT is at its best when the
number of infinite summations can be reduced at most to one.

Let's now turn to the radial integration. Having dropped the
exponential and introduced the infrared cutoff, radial integrands
consist of simple powers. The only difficulty is that, because of
the presence of the min and max functions, the domain of
integration (which is an hypercube of length $R$) has to be split
into $(P-\ell)!$ subdomains, defined by the different orderings of
the radial variables. This number can be large. Of course this is
not a problem if the procedure is automated with the help of a
computer\footnote{In particular, we used the \texttt{Mathematica}
routine for radial integrals described in \cite{SiegGPXT}.}.
Anyway, it is useful to find all the possible symmetries of the
integrand in order to reduce the independent domains of
integrations.

At this point, when the angular and radial integrations have been
performed, the next point is to promote the dimension $D$, or
equivalently $\lambda$, to a complex parameter through the formula
(\ref{lambda}) and then perform the Laurent expansion of the
result around $\varepsilon=0$. If multiple poles are present, we
proceed to the subtraction of subdivergences. These must be
computed within the same renormalization scheme as the original
integral, \emph{i.e.} using GPXT and introducing the same cutoff
procedure for infrared regularization.

The last step is to perform the summations that possibly survived
after the angular integrations. As stated before, finding
analytical results can be very hard, especially when multiple
series are present, and sometimes only a numerical analysis is
possible.

\section{Integrals}\label{integrals}
In the following, we list the results of the integrals relevant
for the computation of maximal reshuffling diagrams of ABJM theory
in the $SU(2)\times SU(2)$ sector (see Figure \ref{supergraphs}).
They are computed via GPXT as described in Appendix \ref{GPXT}. We
denote by $I_j$ the value of the integral. We give here the
$\varepsilon$ expansions up to the order needed for the
computation of the UV divergences of the diagrams in Figure
\ref{supergraphs}. We denote by $R$ the infrared regulator. For
convenience, we leave here the normalization factor
(\ref{normalization}) unexpanded: this will be easily reintroduced
in the subtraction of subdivergences.
\begin{equation}\begin{split}
I_2\;=\;&\graph{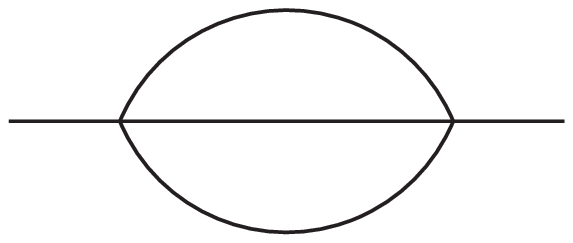}{4}\;=\;N_\lambda(2,3)\,\left(\frac{1}{2\,\varepsilon}+\log R+\varepsilon\,\log^2R+\frac{2}{3}\,\varepsilon^2\,\log^3R\right)\, ,\\
\end{split}\end{equation}
\begin{equation}\begin{split}
I_4\;=\;&\graph{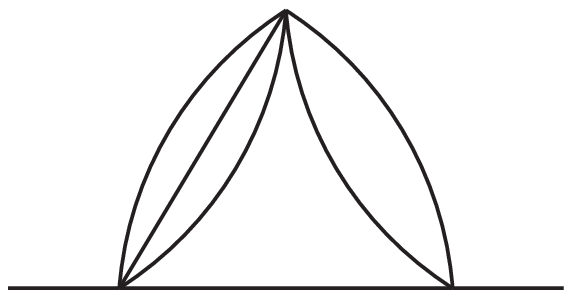}{5}\;=\;N_\lambda(4,6)\,\Big\{\frac{1}{8\,\varepsilon^2}+\frac{1+\log R}{2\,\varepsilon}-1+2\log R+\log^2 R\\
&\phantom{XXXXXXXX}+\frac{2}{3}\,\varepsilon\left(3-6\log R+6\log^2R+2\log^3R\right)\Big\}\, ,\\
\end{split}\end{equation}
\begin{equation}\begin{split}
I_{4a}\;=\;&\graph{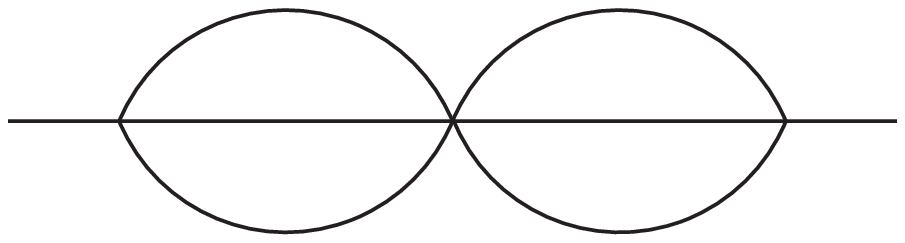}{3.2}\;=\;N_\lambda(4,6)\,\Big(\frac{1}{4\,\varepsilon^2}+\frac{\log R}{\varepsilon}+2\log^2R+\frac{8}{3}\,\varepsilon\log^3R\Big)\, ,\\
\end{split}\end{equation}
\begin{equation}\begin{split}
I_{6}\;=\;&\graph{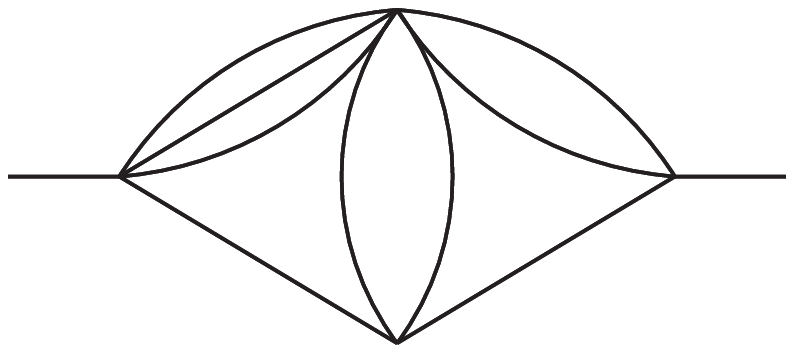}{5}\;=\;N_\lambda(6,9)\Big\{\frac{1}{48\,\varepsilon^3}+\frac{2+\log
R}{8\,\varepsilon^2}+\frac{1}{\varepsilon}\Big(\frac{5}{6}+\frac{3}{2}\log
R+\frac{3}{8}\log^2
R\Big)\\
&\phantom{XXXXXXXXXXX}-\frac{29}{3}+5\log R+\frac{9}{2}\log^2R+\frac{3}{4}\log^3R\Big\}\\
\end{split}\end{equation}
\begin{equation}\begin{split}
I_{6a}\;=\;&\;\;\graph{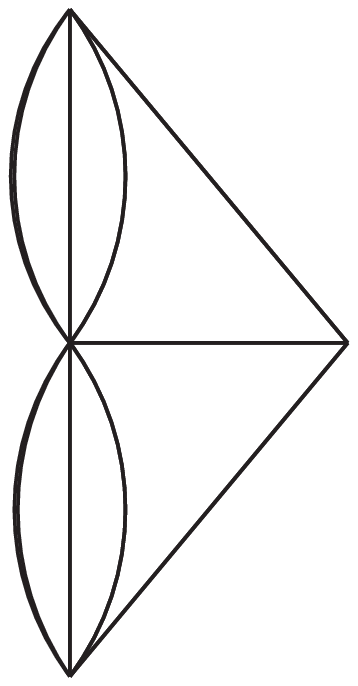}{8}\;\;\;=\;N_\lambda(6,9)\Big\{\frac{1}{24\,\varepsilon^3}+\frac{1}{\varepsilon^2}\,\left(\frac{1}{3}+\frac{1}{4}\log
R\right)+\frac{1}{\varepsilon}\Big(-\frac{4}{3}+2\log
R+\frac{3}{4}\log^2
R\Big)\\
&\phantom{XXXXXXXXX}+8-8\log R+6\log^2R+\frac{3}{2}\log^3R\Big\}\\
\end{split}\end{equation}
\begin{equation}\begin{split}
I_{6b}\;=\;&\;\;\graph{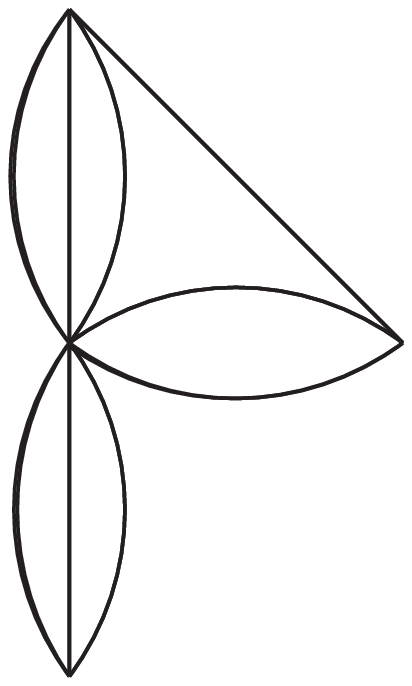}{8}\;\;\;=\;N_\lambda(6,9)\Big(\frac{1}{16\,\varepsilon^3}+\frac{2+3\log
R}{8\,\varepsilon^2}+\frac{-4+12\log R+9\log^2
R}{8\,\varepsilon}\\
&\phantom{XXXXXXXXXX}+1-3\log R+\frac{9}{2}\log^2R+\frac{9}{4}\log^3R\Big)\\
\end{split}\end{equation}
\begin{equation}\begin{split}
I_{6c}\;=\;&\graph{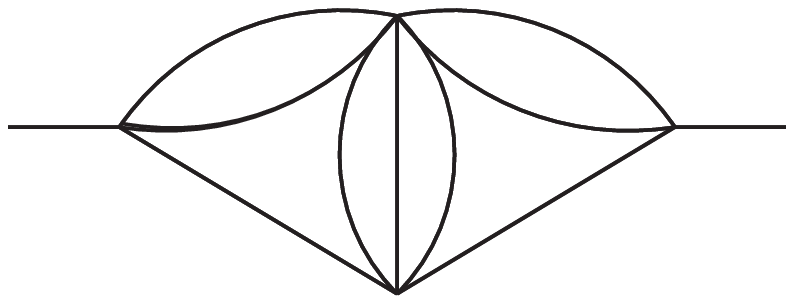}{5}\;=\;N_\lambda(6,9)\Big\{\frac{1}{24\,\varepsilon^3}+\frac{1}{\varepsilon^2}\Big(\frac{1}{6}+\frac{1}{4}\log
R\Big)+\frac{1}{\varepsilon}\Big(\frac{1}{3}+\log R+\frac{3}{4}\log^2R\Big)\\
&\phantom{XXXXXXXXXX}-2+2\log R+3\log^2 R\Big\}\\
\end{split}\end{equation}
\begin{equation}\begin{split}
I_{8a}\;=\;&\;\;\graph{I8a}{8}\;=\;N_\lambda(8,12)\Big\{\frac{1}{192\,\varepsilon^4}+\frac{1}{12\,\varepsilon^3}\Big(1+\frac{1}{2}\log
R\Big)\\
&\phantom{XXXXXX}+\frac{1}{3\,\varepsilon^2}\Big(-\frac{1}{4}+2\log
R+\frac{1}{2}\log^2
R\Big)\\
&\phantom{XXXXXX}+\frac{1}{3\,\varepsilon}\Big(x-2\log R+8\log^2R+\frac{4}{3}\log^3R\Big)\Big\}\, ,\\
\end{split}\end{equation}
\begin{equation}\begin{split}
I_{8b}\;=\;&\;\;\graph{I8b}{5}\;=\;N_\lambda(8,12)\Big\{\frac{5}{384\,\varepsilon^4}+\frac{5}{48\,\varepsilon^3}\Big(1+\log
R\Big)\\
&\phantom{XXXXXXXXX}+\frac{1}{6\,\varepsilon^2}\Big(\frac{1}{2}+5\log
R+\frac{5}{2}\log^2
R\Big)\\
&\phantom{XXXXXXXXX}+\frac{1}{\varepsilon}\Big(-\frac{7}{4}+\frac{2}{3}\log R+\frac{10}{3}\log^2R+\frac{10}{9}\log^3R\Big)\Big\}\\
\end{split}\end{equation}
\begin{equation}\begin{split}
I_{8c}\;=\;&\;\;\graph{I8c}{5}\;=\;N_\lambda(8,12)\Big\{\frac{1}{128\,\varepsilon^4}+\frac{1}{4\,\varepsilon^3}\Big(\frac{1}{3}+\frac{1}{4}\log
R\Big)\\
&\phantom{XXXXXXXXX}+\frac{1}{\varepsilon^2}\Big(\frac{5}{24}+\frac{2}{3}\log
R+\frac{1}{4}\log^2
R\Big)\\
&\phantom{XXXXXXXXX}+\frac{1}{3\,\varepsilon}\Big(\frac{1}{2}+5\log R+8\log^2R+2\log^3R\Big)\Big\}\\
\end{split}\end{equation}
\begin{equation}\begin{split}
I_{8d}\;=\;&\graph{I8d}{5}\;=\;N_\lambda(8,12)\Big\{\frac{1}{128\,\varepsilon^4}+\frac{1}{8\,\varepsilon^3}\Big(1+\frac{1}{2}\log
R\Big)\\
&\phantom{XXXXXXXXX}+\frac{1}{\varepsilon^2}\Big(\frac{1}{8}+\log
R+\frac{1}{4}\log^2 R\Big)\\
&\phantom{XXXXXXXXX}+\frac{1}{\varepsilon}\Big(-3+\log
R+4\log^2R+\frac{2}{3}\log^3R\Big)\Big\}
\end{split}\end{equation}
\begin{equation}\begin{split}
I_{8e}\;=\;&\graph{I8}{5}\;=\;N_\lambda(8,12)\Big\{\frac{1}{384\,\varepsilon^4}+\frac{1}{8\,\varepsilon^3}\Big(\frac{1}{2}+\frac{1}{6}\log
R\Big)\\
&\phantom{XXXXXXXXX}+\frac{1}{\varepsilon^2}\Big(\frac{2}{3}+\frac{1}{2}\log
R+\frac{1}{12}\log^2
R\Big)\\
&\phantom{XXXXXXXXX}+\frac{1}{\varepsilon}\Big(-\frac{7}{12}+\frac{16}{3}\log
R+2\log^2R+\frac{2}{9}\log^3R\Big)\Big\}\\
\end{split}\end{equation}
where, in $I_{8a}$, $x$ is reduced to a multiple series which we
don't show here (see below for a more effective method to compute
the $I_{8a}$ integral). The subtracted integrals are denoted by
$\bar{I}_j$ and are graphically represented by a box around the
corresponding integral. We conveniently factor out
$1/(4\pi)^{\ell}$ from the results. The pole parts of such
integrals are given by:
\begin{equation}\begin{split}
\bar{I}_2\;=\;&\graph{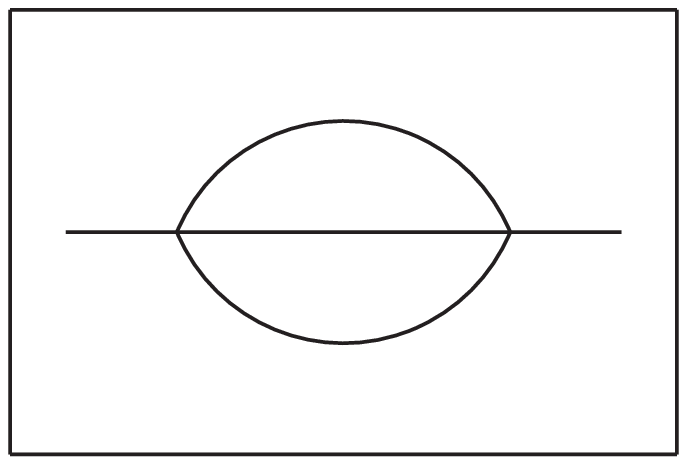}{5}\;=\;\frac{1}{(4\pi)^2}\,\frac{1}{4\,\varepsilon}\\
&\phantom{}\\
\end{split}\end{equation}
\begin{equation}\begin{split}
\bar{I}_4\;=\;&\graph{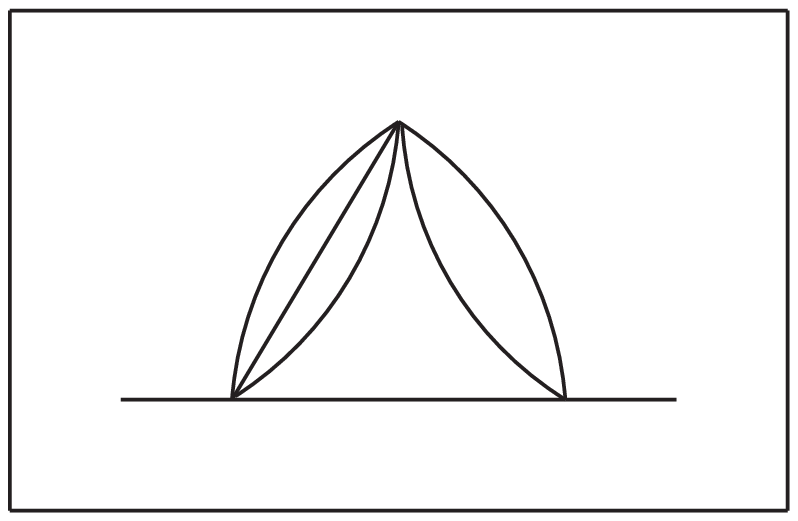}{6}\;=\;\graph{I4}{4}\;-\;\graph{I2box}{6}\graph{I2}{4}\;=\;
\frac{1}{(4\pi)^4}\left(-\frac{1}{32\,\varepsilon^2}+\frac{1}{8\,\varepsilon}\right)\\
&\phantom{}\\
\end{split}\end{equation}
\begin{equation}\begin{split}
\bar{I}_{4a}\;=\;&\graph{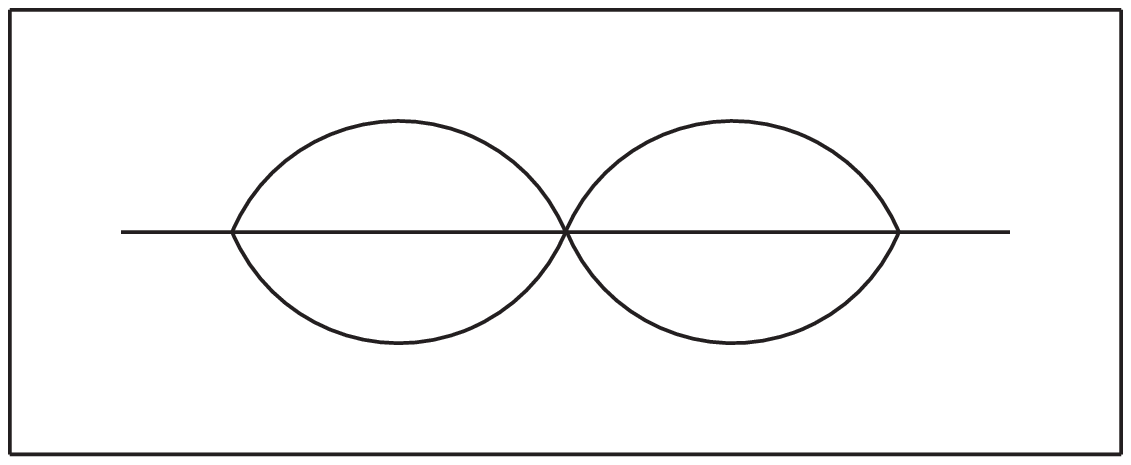}{6}\;=\;\graph{I4a}{4}\;-\;2\;\graph{I2box}{6}\graph{I2}{4}\\
&\phantom{}\\
&\phantom{XXXXXXXXXX}=\;\frac{1}{(4\pi)^4}\left(-\frac{1}{16\,\varepsilon^2}\right)\\
&\phantom{}\\
\end{split}\end{equation}
\begin{equation}\begin{split}
\bar{I}_{6}\;=\;&\graph{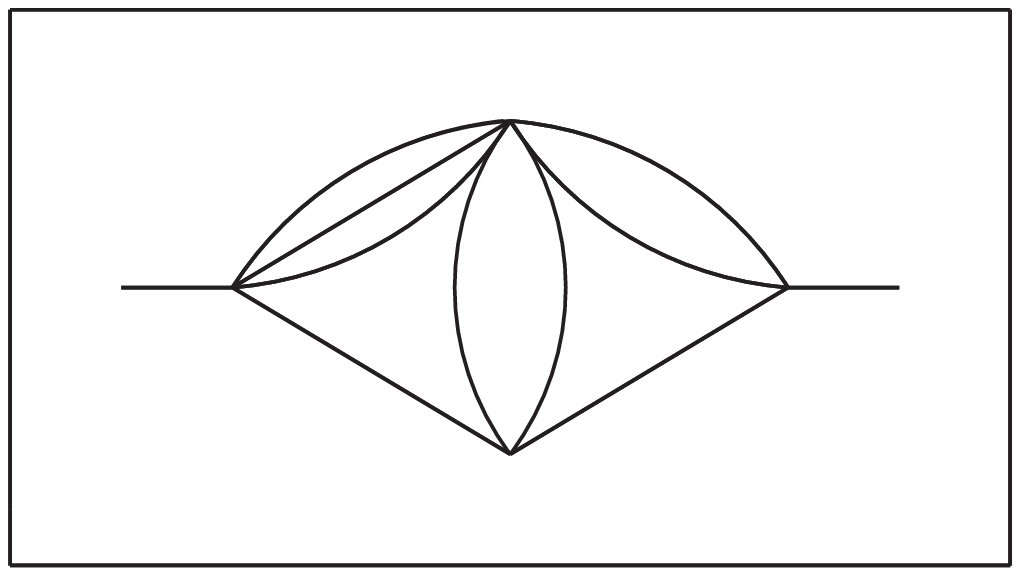}{6}\;=\;\graph{I6}{4}\;-\graph{I2box}{6}\graph{I4}{4}\;-\;\graph{I4box}{5}\graph{I2}{4}\\
&\phantom{}\\
&\phantom{XXXXXXXX}=\;\frac{1}{(4\pi)^6}\left(\frac{1}{384\,\varepsilon^3}-\frac{1}{32\,\varepsilon^2}+\frac{1}{6\,\varepsilon}\right)\\
&\phantom{}\\
\end{split}\end{equation}
\begin{equation}\begin{split}
\bar{I}_{6a}\;=\;&\;\;\graph{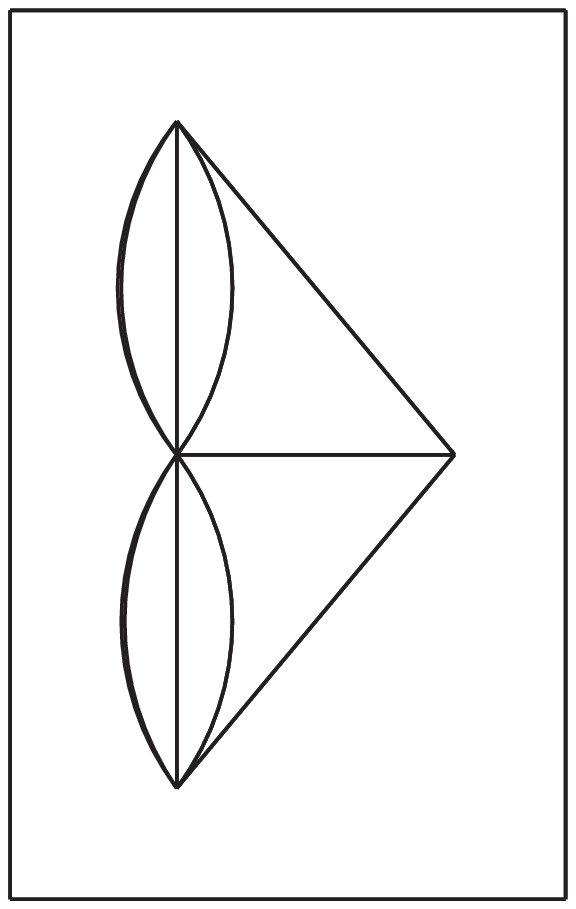}{10}\;\;\;=\;\;\;\graph{I6a}{8}\;\;\;-\;2\;\graph{I2box}{6}\graph{I4}{4}\;-\;\graph{I4abox}{5}\graph{I2}{4}\\
&\phantom{}\\
&\phantom{XXXXXXXX}=\;\frac{1}{(4\pi)^6}\left(\frac{1}{192\,\varepsilon^3}-\frac{1}{48\,\varepsilon^2}-\frac{1}{24\,\varepsilon}\right)\\
&\phantom{}
\end{split}\end{equation}
\begin{equation}\begin{split}
\bar{I}_{6b}\;=\;&\;\;\graph{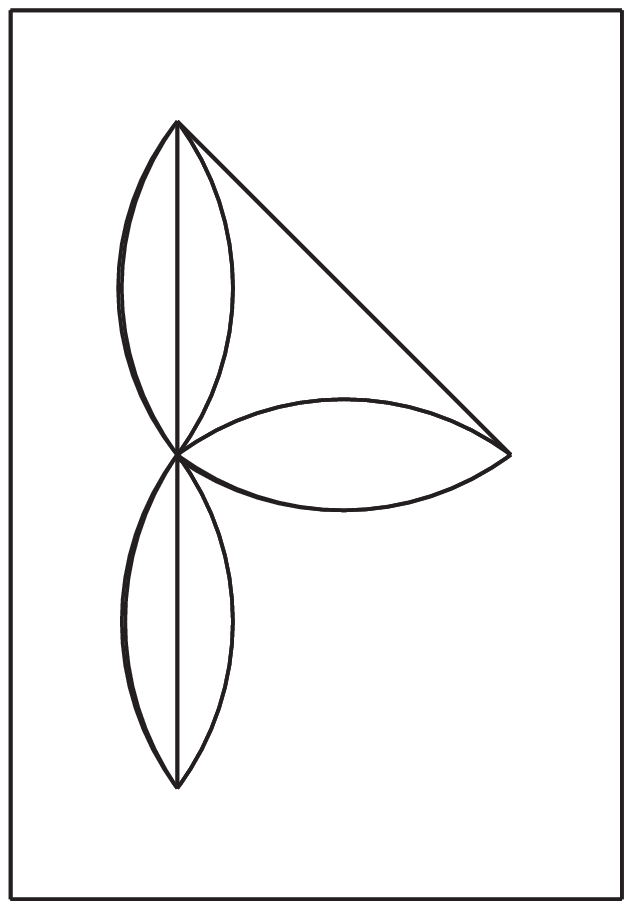}{10}\;\;\;=\;\;\;\graph{I6b}{8}\;\;\;-\;\graph{I2box}{6}\;\left(\graph{I4}{4}\;+\;\graph{I4a}{4}\right)\\
&\phantom{}\\
&\phantom{XXXXXX}\;-\;\graph{I4box}{6}\graph{I2}{4}\;-\;\graph{I4abox}{5}\graph{I2}{4}\\
&\phantom{}\\
&\phantom{XXXXXXXX}=\;\frac{1}{(4\pi)^6}\left(\frac{1}{128\,\varepsilon^3}-\frac{1}{32\,\varepsilon^2}\right)\\
&\phantom{}\\
\end{split}\end{equation}
\begin{equation}\begin{split}
\bar{I}_{6c}\;=\;&\;\;\graph{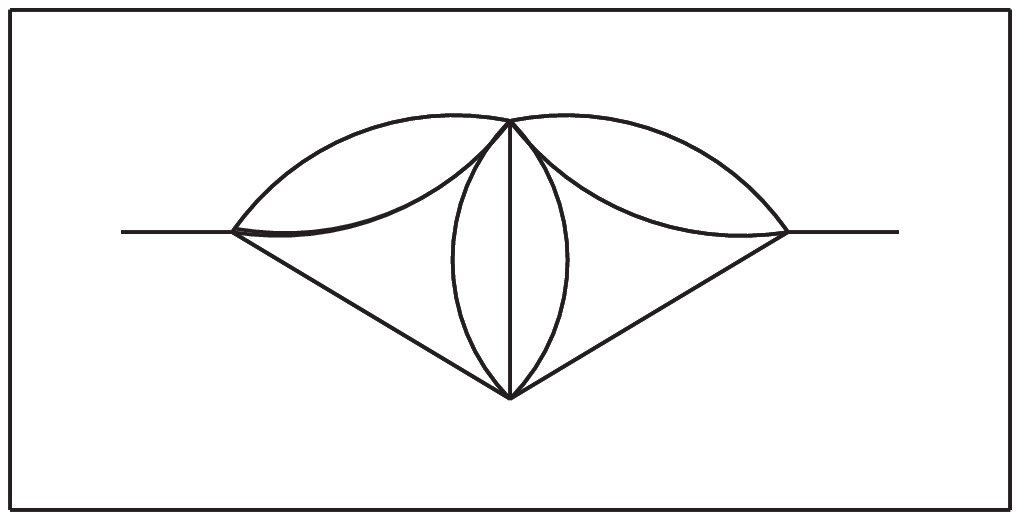}{6}\;\;\;=\;\;\;\graph{I6c}{5}\;\;\;-\;\graph{I2box}{6}\;\graph{I4a}{4}\\
&\phantom{}\\
&\phantom{XXXXXX}\;-\;2\;\graph{I4box}{6}\graph{I2}{4}\\
&\phantom{}\\
&\phantom{XXXXXXXX}=\;\frac{1}{(4\pi)^6}\left(\frac{1}{192\,\varepsilon^3}-\frac{1}{24\,\varepsilon^2}+\frac{1}{24\,\varepsilon}\right)\\
&\phantom{}\\
\end{split}\end{equation}
\begin{equation}\label{I8aGPXT}\begin{split}
\bar{I}_{8a}\;=\;&\graph{I8abox}{10}\;=\;\graph{I8a}{8}\;-\;\graph{I2box}{6}\;\graph{I6a}{8}\\
&\phantom{}\\
&\phantom{XXXXXXXX}-\;2\;\graph{I4box}{6}\graph{I4}{4}\;-\;\graph{I6cbox}{6}\graph{I2}{4}\;\\
&\phantom{}\\
&\phantom{XXXXXXX}=\;\frac{1}{(4\pi)^8}\,\Big\{-\frac{1}{3072\,\varepsilon^4}+\frac{1}{192\,\varepsilon^3}-\frac{1}{48\,\varepsilon^2}+\frac{y}{\varepsilon}\Big\}\,
,\\
\end{split}\end{equation}
\begin{equation}\begin{split}
\bar{I}_{8b}\;=\;&\graph{I8bbox}{6}\;=\;\graph{I8b}{5}\;-\;\graph{I2box}{6}\left(\graph{I6c}{5}+\graph{I6b}{8}\right)\\
&\phantom{}\\
&\phantom{XXXXX}\;-\;\graph{I4box}{6}\graph{I4}{4}\;-\;\left(\graph{I6abox}{8}+\graph{I6bbox}{8}\right)\graph{I2}{4}\\
&\phantom{}\\
\;&\phantom{XXXXXX}-\;\graph{I4abox}{6}\graph{I4a}{4}\\
&\phantom{}\\
&\phantom{XXXXXXXX}=\;\frac{1}{(4\pi)^8}\left(-\frac{5}{6144\,\varepsilon^4}+\frac{5}{768\,\varepsilon^3}-\frac{1}{384\,\varepsilon^2}-\frac{1}{32\,\varepsilon}\right)\\
&\phantom{}\\
\end{split}\end{equation}
\begin{equation}\begin{split}
\bar{I}_{8c}\;=\;&\graph{I8cbox}{6}\;=\;\graph{I8c}{5}\;-\;\graph{I2box}{6}\;\graph{I6b}{8}\\
&\phantom{}\\
&\phantom{XXXXXXXX}-\;\graph{I4box}{6}\left(\graph{I4}{4}+\graph{I4a}{4}\right)\\
&\phantom{}\\
&\phantom{XXXXXXXX}-\;\left(\graph{I6cbox}{6}\;+\;\graph{I6box}{6}\right)\graph{I2}{4}\;\\
&\phantom{}\\
&\phantom{XXXXXXX}=\;\frac{1}{(4\pi)^8}\left(-\frac{1}{2048\,\varepsilon^4}+\frac{1}{128\,\varepsilon^3}-\frac{3}{64\,\varepsilon^2}+\frac{5}{192\,\varepsilon}\right)\\
&\phantom{}\\
\end{split}\end{equation}
\begin{equation}\begin{split}
\bar{I}_{8d}\;=\;&\graph{I8dbox}{7}\;=\;\graph{I8d}{5}\;-\;\graph{I2box}{6}\;\left(\graph{I6}{5}\;+\;\graph{I6a}{8}\right)\\
&\phantom{}\\
&\phantom{XXXXX}\;-\;\left(\graph{I4box}{6}\;+\;\graph{I4abox}{6}\right)\graph{I4}{4}\;-\;\graph{I6bbox}{10}\graph{I2}{4}\\
&\phantom{}\\
&\phantom{XXXXXXXXXXX}=\;\frac{1}{(4\pi)^8}\left(-\frac{1}{2048\,\varepsilon^4}+\frac{1}{192\,\varepsilon^3}-\frac{1}{64\,\varepsilon^2}-\frac{11}{192\,\varepsilon}\right)
\end{split}\end{equation}
\begin{equation}\begin{split}
\bar{I}_{8e}\;=\;&\graph{I8box}{7}\;=\;\graph{I8}{5}\;-\;\graph{I2box}{6}\graph{I6}{5}\\
&\phantom{}\\
&\phantom{XXXXX}\;-\;\graph{I4box}{6}\graph{I4}{4}\;-\;\graph{I6box}{6}\graph{I2}{4}\\
&\phantom{}\\
&\phantom{XXXXXXXXXXX}=\;\frac{1}{(4\pi)^8}\left(-\frac{1}{6144\,\varepsilon^4}+\frac{1}{256\,\varepsilon^3}-\frac{19}{384\,\varepsilon^2}+\frac{5}{16\,\varepsilon}\right)\, .\\
&\phantom{}\\
\end{split}\end{equation}
Regarding $\bar{I}_{8a}$, GPXT allowed us to analytically compute
the higher order poles in $\varepsilon$ and to reduce the first
order pole $y$ to a multiple series. Such a series is not easy to
sum and we choose to compute the integral $\bar{I}_{8a}$ through
Mellin-Barnes technique \cite{Smirnov1,Smirnov2} in order to have
a reliable numerical estimate of its pole
$\bar{I}_{8a}|_{\frac{1}{\varepsilon}}$.

First of all, we contract the bubbles to reduce to the evaluation
of a four-loop integral:
\begin{equation}\label{MBI8a}
I_{8a}\;=\;\graph{I8a}{10}\;=\;G(1,1)^3\,G(1,1/2+\varepsilon)\;\graph{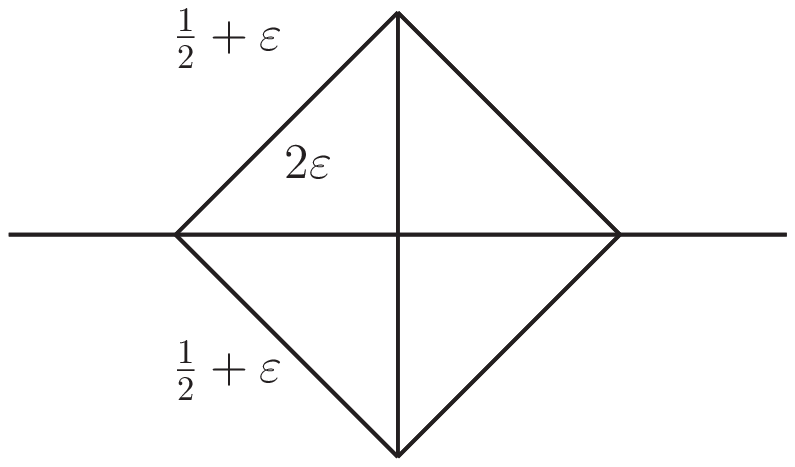}{10}\,
,
\end{equation}
where the ``G-functions'' are defined by
\begin{equation}
G(\alpha,\beta)=\,\frac{\Gamma(\lambda+1-\alpha)\Gamma(\lambda+1-\beta)\Gamma(\alpha+\beta-\lambda-1)}{(4\pi)^{\lambda+1}\Gamma(\alpha)\Gamma(\beta)\Gamma(2\lambda+2-\alpha-\beta)}\,
.
\end{equation}
Since the G-functions in (\ref{MBI8a}) present a simple pole in
$\varepsilon$, the four-loop integral
\begin{equation}
I_{4b}\;\;\;=\;\;\;\graph{I4b}{12}
\end{equation}
has to be computed up to order zero in $\varepsilon$. Let's
introduce the corresponding four-loop master integral with generic
powers of the propagators:
\begin{equation}\begin{split}
J_{4b}(\alpha_1,\ldots,\alpha_8)\;\;\;=\;&\;\;\graph{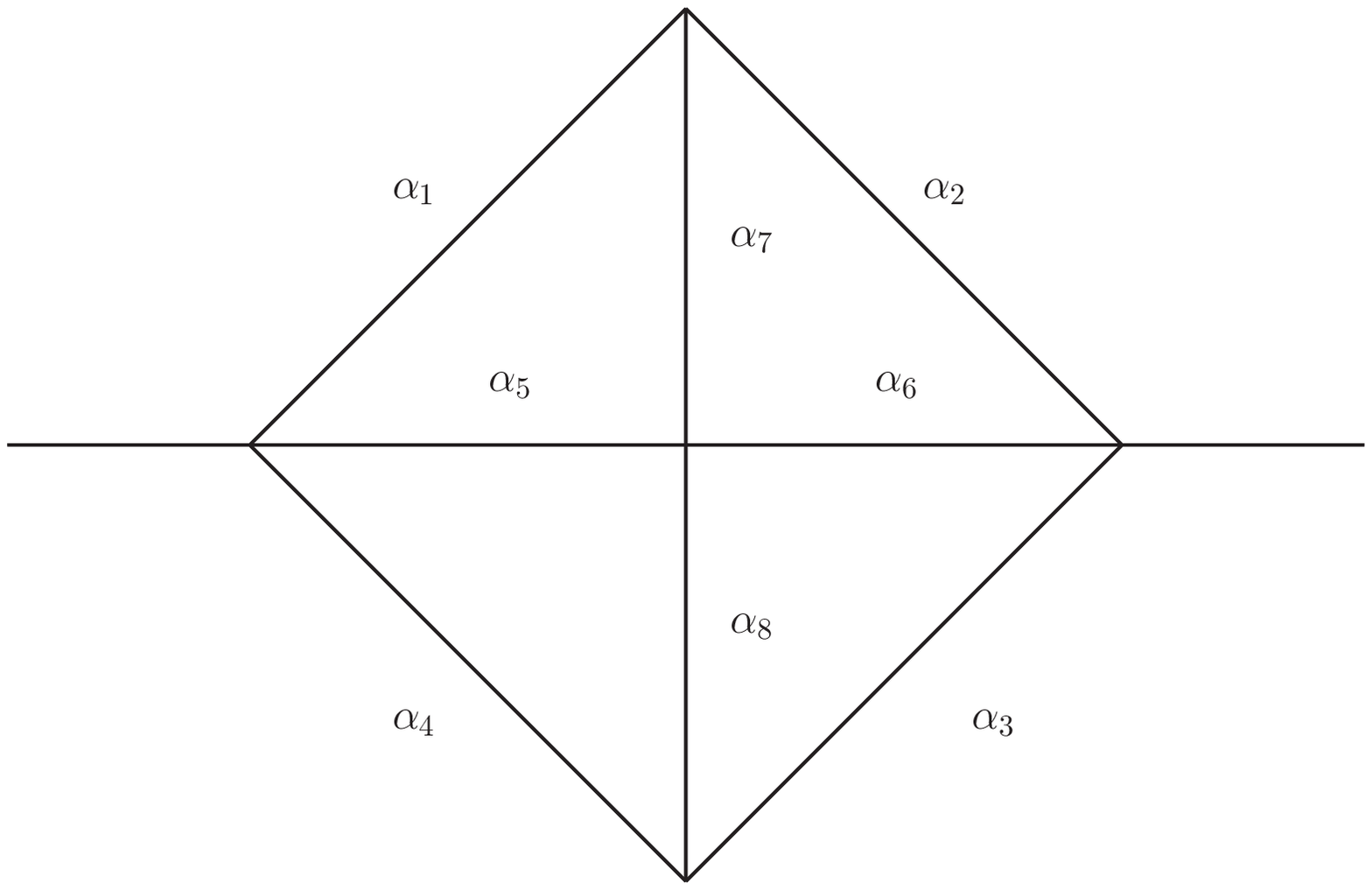}{15}\\
=\;&\frac{1}{(2\pi)^{4D}}\,\int\,\di^D k_1\di^D k_2\di^D k_3\di^D
k_4\,\frac{1}{(k_1^2)^{\alpha_1}(k_2^2)^{\alpha_2}(k_3^2)^{\alpha_3}(k_4^2)^{\alpha_4}}\\
&\;\;\times\frac{1}{[(k_1-k_4-p)^2]^{\alpha_5}[(k_2-k_3-p)^2]^{\alpha_6}[(k_1-k_2)^2]^{\alpha_7}[(k_3-k_4)^2]^{\alpha_8}}\,
.
\end{split}\end{equation}
The Mellin-Barnes representation of this integral and its
analytical continuation in $\varepsilon$ can be obtained with the
help of the \texttt{Mathematica} packages \texttt{AMBRE}
\cite{Gluza:2007rt} and \texttt{MB} \cite{Czakon:2005rk}
respectively.

The \texttt{MB} package also contains a routine suitable for
numerical integration of MB representations. Besides the built-in
\texttt{Mathematica} function \texttt{NIntegrate}, it uses the
\texttt{CUBA} library \cite{Hahn:2004fe} of numerical integration
routines and the CERN libraries \cite{CERN} for the implementation
of gamma and psi functions, in order to prepare \texttt{Fortran}
programs, which are more efficient, in terms of computational
time, for high dimension MB integrals.

A MB representation for the integral $J_{4b}$ is
\begin{equation}\label{MB}\begin{split}
J_{4b}=&\frac{1}{(4\pi)^{\lambda+1}}\,\int_{-i\infty}^{i\infty}\frac{\di
z_1}{2\pi i}\cdots\int_{-i\infty}^{i\infty}\frac{\di z_6}{2\pi
i}\;\frac{\Gamma(\lambda+1-\alpha_{17}-z_1)\Gamma(-z_1)\Gamma(\lambda+1-
\alpha_{15}-z_2)}{\Gamma(\alpha_1)
\Gamma(\alpha_3)\Gamma(\alpha_5)\Gamma(\alpha_6)}\\
&\phantom{}\\
&\times\frac{\Gamma(-z_2)\Gamma(\alpha_1+z_{12})\Gamma(\lambda+1-\alpha_{26}+
z_1-z_3)\Gamma(2\lambda+2-\alpha_{1257}-z_{24})\Gamma(-z_4)}{\Gamma(2\lambda+2-\alpha_{157})
\Gamma(\alpha_7)\Gamma(\alpha_2-z_1)\Gamma(3\lambda+3-\alpha_{12567}-z_2)}\\
&\phantom{}\\
&\times\frac{\Gamma(\alpha_2-z_1+z_{34})
\Gamma(-2\lambda-2+\alpha_{12567}+z_{234})\Gamma(3\lambda+3-\alpha_{1235678}-z_{2345})
}{\Gamma(4\lambda+4-\alpha_{1235678}-z_{24})\Gamma(-2\lambda-2+\alpha_{125678}+z_{234})
\Gamma(\alpha_4-z_5)}\\
&\phantom{}\\
&\times\frac{\Gamma(-z_5)\Gamma(\lambda+1-\alpha_4+z_5)\Gamma(\lambda+1-\alpha_3+
z_3-z_6)\Gamma(4\lambda+4-\alpha_{1235678}-z_{56})}{
\Gamma(5\lambda+5-\alpha_{12345678}-z_6)}\\
&\phantom{}\\
&\times\frac{\Gamma(-z_6)
\Gamma(-4\lambda-4+\alpha_{12345678}+z_6)\Gamma(\alpha_3+z_{56})
\Gamma(-3\lambda-3+\alpha_{1235678}+z_{2456})}{
\Gamma(-3\lambda-3+\alpha_{1235678}+z_{56})}\, ,
\end{split}\end{equation}
where we have denoted
$\alpha_{ijk\cdots}=\alpha_i+\alpha_j+\alpha_k+\cdots$ and
similarly for $z_k$.

We have to compute
\begin{equation}
I_{4b}\;=\;J_{4b}(1/2+\varepsilon,1,1,1/2+\varepsilon,2\varepsilon,1,1,1)\,
.
\end{equation}
Putting this integral into the \texttt{MB} package we can make
the $\varepsilon$-expansion and find the following numerical
result:
\begin{equation}\begin{split}
I_{4b}\,=\,&\frac{2.116213934935895\,10^{-8}}{\varepsilon^3}+
\frac{(8.26225\pm 0.00003)\,10^{-7}}{\varepsilon^2}\\
&\phantom{}\\
&+\frac{(0.0000177432\pm
0.0000000003)}{\varepsilon}+(0.0002705914\pm 0.0000000003)\, .
\end{split}\end{equation}
\\We then insert this result in (\ref{MBI8a}), make the
$\varepsilon$-expansion and subtract the subdivergences, to obtain
the following numerical result:
\begin{equation}\begin{split}
\bar{I}_{8a}\,=\,&-\frac{5.2348\ldots\,10^{-13}}{\varepsilon^4}+
\frac{8.37567\ldots\,10^{-12}}{\varepsilon^3}-\frac{3.35028\ldots\,10^{-11}}{\varepsilon^2}
-\frac{9.63613\ldots\,10^{-12}}{\varepsilon}\, .
\end{split}\end{equation}
Comparing this result with (\ref{I8aGPXT}) we see that the poles
of second, third and fourth order perfectly agree. We recall that
these poles were computed analytically via GPXT. In addition, we
now have the numerical result for the first order pole:
\begin{equation}\label{I8anum}
\bar{I}_{8a}|_{1/\varepsilon}\;=\;\frac{1}{(4\pi)^8}\,(-0.0059921\pm
0.0000008)\, .
\end{equation}
We now want to extract the analytical result from this number.
In section \ref{BA} we argued that the parameter $\beta$ of
(\ref{dilatationmr}) is purely transcendental and that it can be a
rational combination of just two transcendental constants, namely
$\pi^3$ and $\zeta(3)$. Since the only transcendental contribution
to $\beta$ can come from the $\frac{1}{\varepsilon}$ pole of
$\bar{I}_{8a}$, such pole should be a rational combination of the
constants 1, $\pi^3$, $\zeta(3)$. In order to extract the
coefficients of this generic linear combination from the numerical
value (\ref{I8anum}) we use the \texttt{Mathematica}
implementation \cite{PSLQnb} of the PSLQ algorithm\footnote{A
similar strategy has been used \emph{e.g.} in \cite{Bak:2009tq}}
\cite{PSLQ} (see also \cite{PSLQ2}). This is a powerful integer
relation detection algorithm: given a vector $x=(x_1,\ldots,x_n)$
of real or complex numbers and a precision $10^{-p}$, PSLQ looks
for a vector $a=(a_1,\ldots,a_n)$, if exists, of integers $a_i$,
not all zero, such that
\begin{equation}
a_1 x_1+\cdots+a_n x_n=0\, .
\end{equation}
In our case, the vector $x$ contains the numerical value
in $\bar{I}_{8a}\mid_{1/\varepsilon}$ and the transcendental
constants it should be fitted with:
\begin{equation}
x=\big(-0.0059921,1,\pi^3,\zeta(3)\big)\, .
\end{equation}
Application of PSLQ algorithm with precision $10^{-7}$, which is
the same within which the numerical value of
$\bar{I}_{8a}\mid_{1/\varepsilon}$ has been computed, gives the
solution
\begin{equation}
a=(-32,-5,0,4)\, .
\end{equation}
It means that the exact value of the $\frac{1}{\varepsilon}$
pole of the $\bar{I}_{8a}$ integral should be
\begin{equation}
\bar{I}_{8a}\mid_{1/\varepsilon}\;=\;\frac{1}{(4\pi)^8}\,\left(-\frac{5}{32}+\frac{1}{8}\,\zeta(3)\right)\,
.
\end{equation}
We note that the precision obtained for the result  (\ref{I8anum}) comes from the numerical integration of the highest dimensional MB integrals performed by the \texttt{CUBA} library  \cite{Hahn:2004fe} routines used by the \texttt{MB} package. Obtaining a better precision would require a balanced modification of the library working parameters.



\end{document}